\documentclass{ieeeaccess}

\usepackage{cite}
\usepackage{amsmath,amssymb,amsfonts}
\usepackage{graphicx}
\usepackage{textcomp}
\usepackage{float}
\usepackage{placeins}
\usepackage{subfig}
\usepackage{dblfloatfix}
\usepackage{caption}
\usepackage{array}
\usepackage{flushend}
\usepackage{comment}
\usepackage{url}
\usepackage{xcolor}
\usepackage{etoolbox}

\def\BibTeX{{\rm B\kern-.05em{\sc i\kern-.025em b}\kern-.08em
T\kern-.1667em\lower.7ex\hbox{E}\kern-.125emX}}

\begin{document}

\history{}
\doi{}

\title{Direction-Dependent Path Loss Modeling in Olive Orchards for Precision Agriculture}
\author{
\uppercase{Mohammad Rowhani Sistani}\authorrefmark{1}\authorrefmark{2}\authorrefmark{3},
\uppercase{Katarzyna Kosek-Szott}\authorrefmark{4},
\uppercase{Pierluigi Gallo}\authorrefmark{1}\authorrefmark{5}
}

\address[1]{Department of Engineering, University of Palermo (UNIPA), Palermo, Italy}
\address[2]{School of Advanced Studies, University of Camerino (UNICAM), Camerino, Italy}
\address[3]{SEEDS s.r.l., Italy}
\address[4]{Department of Telecommunications, AGH University of Science and Technology, Krakow, Poland}
\address[5]{CNIT -- Consorzio Nazionale Interuniversitario per le Telecomunicazioni, Italy}

\corresp{Corresponding author: Mohammad Rowhani Sistani (e-mail: mohammad.rowhanisistani@unipa.it)}

\tfootnote{This work was supported in part by the Smart Olive Oil Traceability (SMOOL) Project.}

\markboth
{Rowhani Sistani \headeretal: Direction-Dependent Path Loss Modeling in Olive Orchards}
{Rowhani Sistani \headeretal: Direction-Dependent Path Loss Modeling in Olive Orchards}

\begin{abstract}
Wireless links deployed in orchards often exhibit significant variability in the strength of the received signal that is not adequately captured by classical distance-based propagation models. 
In row-structured olive groves, signal attenuation differs markedly between along-row and cross-row propagation directions, leading to discrepancies when using omnidirectional propagation assumptions such as those adopted in the free-space path loss (FSPL) model or International Telecommunication Union Radiocommunication Sector (ITU-R) vegetation loss formulations. This paper proposes a topology-based propagation model that explicitly accounts for orchard layout and the relative positions of radio devices within the plantation structure. 
Experimental validation was conducted using Long Range (LoRa) technology operating at 868 MHz, and the results were compared with established models from the literature and with the proposed two-dimensional model.
The proposed approach achieves a closer fit to measured received signal strength indicator (RSSI) data than conventional models, providing a more reliable basis for link budgeting and network planning in structured agricultural environments.
\end{abstract}

\begin{keywords}
LoRa, path loss modeling, olive orchards, wireless propagation, agricultural IoT, direction-dependent attenuation
\end{keywords}

\titlepgskip=-15pt
\maketitle

\section{Introduction}

Modules based on Long Range (LoRa) technology are becoming a practical option for transmitting data from field activities in commercial orchards. In olive production, for example, low-power nodes can be attached to harvesting bins while a mobile gateway can be installed on a tractor, which moves around the field and collects data as work progresses. For these deployments, to remain usable, path-loss estimates that reflect how radio signals actually propagate within the planted area are needed. This is especially important in order to ensure a complete coverage, e.g., in today's orchards and structured plantation patterns.

The conventional propagation models used for LoRa, described in Section \ref{releatedworks}, are typically calibrated in open fields or generic foliage scenarios. They provide a useful approximation, but describe the environment as horizontally homogeneous and express attenuation as a function of a single distance variable \cite{Myagmardulam2021}.
In a real olive grove, the situation is different; trees are planted on a grid, canopies occupy a sizable portion of the inter-row space, and the tractor follows fixed corridors between them. Under these conditions, path loss depends not only on how far the receiver is from the transmitter but also on how the nodes are oriented with respect to the rows and how much canopy it crosses.

Existing studies on radio propagation in plantations and orchards confirm that vegetation and geometry matter, yet most of the results are reported as one-dimensional distance fits or as case-specific maps \cite{Anzum2022}, \cite{Quispetupa2022}.
In this study, we address this gap by analyzing 868 MHz LoRa modules in an olive orchard in Palermo, Sicily, under a realistic operating scenario with a tractor-mounted gateway and bin-mounted sensor nodes. %To complement the field campaign, we use the CupCarbon simulator to encode the actual orchard layout and measurement geometry, providing a controlled setting to examine the directional effects on LoRa propagation \cite{Mehdi2014CupCarbon}.

Additionally, we propose a two-dimensional (2-D) direction-dependent path-loss model that operates directly on the orchard grid. Unlike distance-only formulations, this approach distinguishes between links with similar Euclidean separation but different orientations across the tree rows. Compared with representative isotropic baselines, it 
reproduces the measured signal distribution over the field more accurately and offers a practical tool for planning LoRa coverage in structured olive groves. The main contributions of this study are 
the experimental characterization of LoRa propagation in a real olive orchard, the formulation of a geometry-aware path-loss model, and its comparison with representative baseline models used in vegetation-affected environments.

\section{Related Works}
\label{releatedworks}

In the realm of the Internet of Things (IoT), the communication side, making sure stable connections in agricultural environments, presents unique challenges. Although LoRa technology is designed for low-power, wide-area communication, its performance is highly sensitive to environmental conditions such as dense vegetation, uneven terrain, and low antenna placement. These factors can significantly attenuate signal strength, resulting in packet loss and unreliable data transfer.
Therefore, reliable communication in such scenarios is key, as intermittent connectivity can lead to delays in data reporting or missed activation signals. Making certain efficient and stable data transmission requires not only low-power design, but also context-aware deployment and adaptive tuning of transmission parameters. %However, in real-world implementations, especially in agricultural or forested.

In the literature, it has been shown that a LoRa node could achieve a range of 5 km under clear line-of-sight (LoS) conditions, while the range decreased to approximately 1 km when the link was blocked by obstructions \cite{AyodejiAgbolade2023}. This underscores the critical impact of such obstacles, and limited propagation in non-line-of-sight (NLoS) scenarios, which leads to severe signal attenuation. Field experiments in fruit orchards have shown that dense vegetation can greatly limit communication distance. In one case, LoRa transmissions with a moderate spreading factor (SF) became undetectable above about 40 m under a thick canopy \cite{Ahmed2024}. These studies show that heavy vegetation cover can significantly shorten the effective range of LoRa compared to open field conditions \cite{Anzum2022}.

\subsection{Generic Path Loss Models}

For baseline estimation, the \textbf{free-space path loss (FSPL)} model is often used to approximate LoRa signal attenuation under ideal conditions \cite{Shamshiri2024}, but in real orchard deployments, these assumptions do not hold true \cite{Mahjoub2024}: foliage and trunk density produce direction-dependent losses that FSPL cannot resolve. 
The FSPL is given by:
\begin{equation}
PL_{\mathrm{FS}}(dB) = -27.55 + 20\log_{10}\!\big(f_{\mathrm{MHz}}\big) + 20\log_{10}\!\big(d_{\mathrm{m}}\big)
\end{equation}
where $f_{\mathrm{MHz}}=868$ and $d_{\mathrm{m}}$ is the distance given in meters, and -27.55 is the constant term for the frequency given in MHz and the distance given in meters. This simple formula provides a convenient reference to compare with real-world path loss measurements in LoRa deployments \cite{FSPL}, \cite{1697062}, \cite{Sulyman2016}.

% According to our focus in this study, in a dense forest area near Hingna village, Nagpur district, Maharashtra, India, the main suppressing factors are physical obstacles, such as trees, dense undergrowth, and uneven terrain, which can cause significant signal interruptions. These obstacles scatter or absorb radio waves, especially at low antenna heights, leading to increased path loss. As a result, packets may not reach their destination, resulting in missed data, delayed communication, or even total loss of connectivity over longer distances \cite{Pradhan2023}. This makes it difficult to maintain stable LoRa links in plantations or forests without flexible solutions.

In practical outdoor deployments, free-space propagation
alone is not sufficient to describe signal attenuation.
When vegetation is present along the propagation path,
an additional excess loss is commonly introduced.
In such cases, the total path loss can be expressed as the
sum of the free-space component and a vegetation-related
attenuation term, as defined in Recommendation P.833 of the International Telecommunication Union Radiocommunication Sector (ITU-R) \cite{RadiocommunicationBureau2021}:

\begin{equation}
PL_{\text{total}}(d) = PL_{\text{FSPL}}(d) + L_{\text{ITU-R}}(f,d)
\end{equation}

On the other side, \textbf{the multi-wall (MW) model} \cite{Azevedo2024}, \cite{Lott2001}, \cite{Anzum2022}, \cite{Sulaiman2012}, extends the log-distance path loss by adding the following excess attenuation of discrete obstacles along the link:

\begin{equation}
PL_{\mathrm{MW}}(d) = PL(d_0) + 20\log_{10}\!({d}) + \sum_{i=1}^{N} L_{w_i}.
\label{eq:mw_classic}
\end{equation}

where $PL_{\mathrm{MW}}(d)$ is the total path loss at distance $d$, $PL(d_0)$ is the reference path loss evaluated at the reference distance that is considered $d_0$ is $1$~m, and $20\log_{10}(d)$ represents the distance-dependent free-space attenuation. The term $\sum_{i=1}^{N} L_{w_i}$ accounts for the cumulative excess loss introduced by $N$ discrete obstacles intersecting the direct propagation path, where $L_{w_i}$ denotes the attenuation associated with the $i$-th wall or floor. The baseline of the MW model is primarily intended for indoor environments, where walls and floors constitute the dominant sources of additional attenuation beyond free-space propagation. These elements are treated as discrete, homogeneous obstructions, whose individual losses are summed along the transmitter–receiver link.

% Following the excess attenuation practice where vegetation loss is added on top of a free-space baseline~\cite{Quispetupa2022} we decompose the measured attenuation into a free-space component and an excess term.
% Using
% \[
% Y(d) = P_{tx} - \mathrm{RSSI}(d)
% \]
% and
% \[
% PL_{\mathrm{FS}}(d) = 32.45 + 20\log_{10}\!\big(f_{\mathrm{MHz}}\big) + 20\log_{10}\!\big(d_{\mathrm{km}}\big),
% \]
% the excess loss is defined, as in~\cite{Quispetupa2022}, by
% \begin{equation}
% Z(d) \;\triangleq\; Y(d) - PL_{\mathrm{FS}}(d).
% \end{equation}
% Empirically, $Z(d)$ follows a logarithmic trend; the best-fit curve is
% \begin{equation}
% Z(d) = a\,\ln d + b, \qquad (a,b) = (4.52,\;24.16),
% \end{equation}
% where $\ln(\cdot)$ denotes the natural logarithm and $d$ is expressed in meters.

\subsection{Path Loss Models in Agricultural Orchards}

Recent studies on wireless propagation in agricultural orchards have shown that tree morphology can substantially affect radio performance \cite{Rao2016}. Measurements carried out in date palm plantations with ZigBee transceivers mounted close to the trunks revealed that the received signal strength depends strongly on antenna height and relative orientation between the transmitter and the tree rows. Directional signals along the rows maintained higher strength and lower path loss than those crossing the rows, indicating direction-dependent propagation in structured orchards. However, the analysis remained descriptive and relied on a one-dimensional log-distance fit, no directional coefficients or two-dimensional path loss formulation was introduced. This leaves a clear gap, despite the empirical evidence of directionality.
%and common empirical models (log-distance, MED, ITU-R, COST-235) effectively assume isotropy. 
We address this gap with a 2-D direction-dependent model, described in Section~\ref{sec:proposed_model}, which parameterizes attenuation separately along columns and rows.

The \textbf{ITU-R vegetation model} provides an experimental baseline for signal attenuation through foliage, widely used to assess propagation in forested or semi-urban areas \cite{RadiocommunicationBureau2021}. It assumes homogeneous vegetation and moderate canopy density, which simplifies modeling at the cost of realism. The additional path loss due to the trees is given by the following formula:
%. The ITU-R model is given by:

\begin{equation}
L_{\text{ITU-R}}\,(\mathrm{dB}) = 0.2 \cdot f_{\mathrm{MHz}}^{0.3} \cdot d_{\mathrm{m}}^{0.6}
\end{equation}

This simplified form, often referred to as an ITU-R–based vegetation model, follows the general structure of ITU-R P.833 but uses coefficients adapted from field data reported in recent studies \cite{Phaiboon2024, Mahjoub2024, Azevedo2024}. We adopt this term because its reported validity range ($200$\,MHz–$95$\,GHz, $d<400$\,m) covers our $868$\,MHz orchard links over tens of meters and provides a clean additive “excess-loss” reference for foliage environments \cite{Mahjoub2024}. However, in structured orchards, tree spacing and trunk periodicity produce direction-dependent propagation paths that the ITU-R formulation cannot fully capture.

A specialization of the multi-wall model in \cite{Anzum2022}, the \textbf{plantation multi-wall (PMW)} considers row-structured trees as discrete obstacles causing additional path loss added to a one-slope baseline:
\begin{equation}\label{eq:mw-olive}
PL_{\mathrm{PMW}}(d_m) \;=\; PL_0 \;+\; 10\,n\,\log_{10} d_m \;+\; L_{\mathrm{canopy}}\, N_{\mathrm{can}}(d_m)
\end{equation}
where $N_{\mathrm{can}}(d_m)$ is the counted canopy crossings along the transmitter--receiver path and $L_{\mathrm{canopy}}$ is the empirical per-canopy loss. The baseline terms of this model are obtained from short-range calibration consistent with the one-slope derivation used by the multi-wall model:
% \begin{subequations}\label{eq:pl0-n}
% \begin{align}
% PL_0 &= 20\log_{10}\!\big(f_{\mathrm{MHz}}\big) - 27.55, \qquad d_0=1~\mathrm{m}, \label{eq:pl0-bca}\\
% n &= \frac{P_r(d_0)-P_r(d)}{10\log_{10}(d/d_0)}, \qquad d_0=1~\mathrm{m}, \label{eq:n-bca}
% \end{align}
% \end{subequations}
$
PL_0 = 20\log_{10}\!\big(f_{\mathrm{MHz}}\big) - 27.55$,
$n = \frac{P_r(d_0)-P_r(d)}{10\log_{10}(d/d_0)}$, and
$d_0=1~\mathrm{m}$.

This model considers the number of intercepted canopies but does not relate it with the orchard layout.
We adopt the PMW model as a baseline and compare it with our proposed model (PM), described in Section~\ref{sec:proposed_model}.

Recent work has addressed LoRa propagation modeling specifically for structured orchard environments. In the \textbf{FLog model} \cite{Yang2023}, a three-dimensional approach was proposed for the shadowing effect of tree canopies and the ground by combining the intrinsic path-loss exponents of different air, foliage, and soil within the First Fresnel Zone (FFZ). By numerically sampling the FFZ, the model estimates the fraction of each medium and integrates them into a weighted Log-Normal Shadowing formulation, validated in almond and walnut orchards.
%showed a 42\%. reduction in path-loss estimation error compared to the conventional log-distance model.
However, while FLog successfully incorporates volumetric shadowing, it remains omnidirectional in nature, the model assumes uniform propagation in all horizontal directions. %This overlooks the pronounced directional attenuation observed between the along-row and across-row orientations, a structural feature of many orchards.
% consider we didnt and also we couldnt due paramters test this model. 
A specific exponent is defined for the
classical log–normal model:
\begin{equation}
PL_{\mathrm{FLog}}(d_m)
= PL(d_0) + 10\, n_{\mathrm{FLog}}
\log_{10}\!\left(\frac{d_m}{d_0}\right) + X_{\sigma},
\label{eq:flog_pl}
\end{equation}
where $PL(d_0)$ is the reference loss at distance $d_0$ and $X_{\sigma}$ a Gaussian random variable that accounts for shadowing, 
%\begin{equation}
$
n_{\mathrm{FLog}} =
P_{\mathrm{open}} \alpha +
P_{\mathrm{foliage}} \beta +
P_{\mathrm{ground}} \gamma ,
\label{eq:flog_n}
$
%\end{equation}
where $P_{\mathrm{open}}$, $P_{\mathrm{foliage}}$, and $P_{\mathrm{ground}}$ denote the
fractions of the FFZ occupied by free space, canopy, and ground, respectively,
and $\alpha$, $\beta$, $\gamma$ are their intrinsic path–loss exponents
obtained via least–squares fitting.

Additionally, recent wide-band field measurements in row-structured crops have quantified direction-dependent loss. In a sugarcane field at 2.45 GHz excess attenuation increased with the number of rows, intersecting the line of sight; \textbf{model selection based on the Akaike Information Criterion (AIC)} and its small-sample correction (AICc) favored a ridge-count vegetation obstruction formulation \cite{srisooksai2018radio}. The presented evidence supports direction-dependent propagation (along-row versus across-row). However, because the model in \cite{srisooksai2018radio} is tied to the ridge counts and local conditions, its transferability and map-scale planning utility are limited. This context motivates our explicit two-dimensional, direction-aware path-loss surface with separate parameters for along-row and inter-row directions. This reflects the fact that attenuation in olive orchards differs significantly from that reported in a tall, densely planted sugarcane field.

% Another work has underscored how environmental geometry governs path-loss behavior. In dense urban canyons, for example, 28 GHz measurements paired with LiDAR and mesh-derived features have enabled machine learning models to capture street-specific variations, reflecting the strong contrast between streets and highway links \cite{Gupta2022}. The same directional confinement arises in row-structured orchards, where foliage alignment and inter-row gaps shape propagation in comparable ways. This connection suggests that an explicit two-dimensional, direction-aware path-loss formulation for sub-GHz orchard settings can complement data-driven urban approaches by describing attenuation separately along and across the rows.

Directional behavior in small-cell networks has been captured by dual-directional path-loss formulations that pair distance with angular dependence \cite{Chen2017}. In these models, the line-of-sight probability varies with direction due to direction-dependent blockage from randomly oriented obstacles, so coverage and rate depend on propagation direction and range. Although developed for urban geometries, the same principle motivates direction-aware modeling in structured orchards, where row alignment and foliage density induce comparable anisotropy.

In summary, the evidence points to a clear need for propagation models that reflect the geometry and materials of working farms, row structure, foliage density, and terrain, rather than relying on distance-only formulas. We follow this path and propose a new 2-D direction-dependent model for olive orchards, described in Section~\ref{sec:proposed_model}. A concise overview of previous work and the remaining gaps in directionality and 2-D modeling is reported in Table~\ref{tab:summarymissing}.
None of the previous models takes care about the regular structure of the orchard considering the distance among rows, columns, and the number of rows and columns between the transmitter and the receiver.

\begin{table}[h]
\centering
\scriptsize
\setlength{\tabcolsep}{1.2pt}
\renewcommand{\arraystretch}{1}
\caption{Propagation models for orchards and gaps.}
\label{tab:summarymissing}
\begin{tabular}{
m{0.1\linewidth}
m{0.1\linewidth}
m{0.18\linewidth} 
%m{0.13\linewidth}
m{0.55\linewidth}
}
\hline
\textbf{Key} & \textbf{Year} & \textbf{Directionality} 
%& \textbf{2D Model ($\Delta x$, $\Delta y$)} 
& \textbf{Notes / Model Focus} \\
\hline

\cite{Lott2001} & 2001 & $\sim$ 
%& $\times$ 
& Multi\textendash Wall\textendash and\textendash Floor (indoor); obstacle penalties, not foliage\textendash row anisotropy. \\
\hline

\cite{Anzum2022} & 2022 & $\sim$ %& $\times$ 
& Foliage\textendash loss Multi\textendash Wall model (palm plantation, 433\,MHz); row structure treated uniformly. \\
\hline
\cite{Yang2023}&2023 & $\sim$ 
%& $\times$ 
& FLog (3D Fresnel\textendash based lognormal) orchard model; horizontally isotropic approximation. \\
\hline
\cite{Quispetupa2022} &2022 & {$\checkmark$} 
%& {$\times$} 
& {Empirical LoRa RSSI characterization in olive orchards; introduces }$\boldsymbol{Z(x)}${ excess\textendash loss term.} \\
\hline

%\cite{Quispetupa2022}&2022 
%& $\times$ 
%& $\times$ & Experimental LoRa RSSI in olive orchards; empirical only, no 2D model. \\
%\hline

%\cite{Sulyman2016}&2016 & %$\checkmark$ 
%& $\times$ 
%& Directional mmWave path\textendash loss (urban); shows angular dependence, not LPWAN/agriculture. \\
%\hline
\cite{Azevedo2024}&2024
& $\times$ & Critical review of LoRa propagation models; no orchard-specific direction-dependent model proposed. \\
\hline

\textbf{This study} & 2025 
%& \textbf{$\checkmark$} 
& \textbf{$\checkmark$} & \textbf{2-D direction-dependent model on }$\boldsymbol{\Delta x}$\textbf{ and }$\boldsymbol{\Delta y}$\textbf{; distinguishes across cols/rows via traversed canopy thickness.} \\
\hline

\end{tabular}
\end{table}

\begin{table*}[h!]
\centering
\caption{Olive orchard planting systems depending on spacing, density, and canopy form across traditional, high-density, and super-high-density layouts \cite{Connor2014,Diez2016,camposeo}.}
\label{tab:olive-systems}
\renewcommand{\arraystretch}{1.15}
\setlength{\tabcolsep}{6pt}
\begin{tabular}{lccc}
\hline
\textbf{System Type} & \textbf{Row $\times$ Tree Spacing (m)} & \textbf{Density (trees ha$^{-1}$)} & \textbf{Canopy / Row Structure} \\
\hline
Traditional & up to $10 \times 10$ & $\approx 100$–$150$ & Isolated crowns; open inter-row spacing \\
Intensive / High-Density (HD) & $5.7 \times 2.25$ to $3.55 \times 1.25$ & $780$–$2254$ & Semi-continuous rows; controlled canopy volume \\
Super-High-Density (SHD) & $4 \times 1.5$ & $\approx 1667$ & Hedgerow-type continuous foliage walls \\
\hline
\end{tabular}
\end{table*}

\begin{comment}
   
\noindent
{\setlength{\fboxsep}{6pt}%
\colorbox{yellow!35}{%
\parbox{\dimexpr\linewidth-2\fboxsep-2\fboxrule\relax}{%
Taken together, the existing literature provides useful references for distance based attenuation, canopy excess loss, and volumetric or direction dependent propagation effects. However, these approaches do not simultaneously capture the regular orchard grid, the row and column dependent propagation geometry, and the resulting spatial RSSI distribution over the plantation. This is the point addressed by the proposed model in the present work.%
}}}
 
\end{comment}

Taken together, the existing literature provides useful references for distance based attenuation, canopy excess loss, and volumetric or direction dependent propagation effects. However, these approaches do not simultaneously capture the regular orchard grid, the row and column dependent propagation geometry, and the resulting spatial received signal strength indicator (RSSI) distribution over the plantation. This is the point addressed by the proposed model in the present work. Machine-learning and Bayesian approaches have also been investigated for uncertainty-aware modeling in sensing and communication-related domains, and may represent a relevant direction for future data-driven extensions of geometry-aware propagation frameworks.

\section{The Olive Orchard Context}

%General system description. The bin- and tractor-related ideas should be explained here.
We assume grid-shaped olive orchards and a transmitter--receiver (Tx/Rx) channel between a mobile gateway mounted on the tractor and a set of compact sensor units integrated into the harvesting bins. The design follows the practical workflow of olive collection, where the tractor naturally moves along the inter-row corridors and the bins remain distributed around the trees during picking activities. This arrangement creates effective communication, with the mobile gateway acting as the receiver and the bin-mounted devices serving as low-power transmitters.

Each bin unit contains a small LoRa transceiver. During normal field operation, the devices are expected to report basic indicators such as location or temperature and other data of harvesting procedures; only the radio link itself is relevant for the present study. The tractor-mounted gateway, positioned approximately at transmitter height, provides a stable reception point in the experimental measurements.

The interaction between these two components, the mobile gateway and the bin nodes, defines the operational setting in which our propagation data were collected. This system configuration mirrors a realistic deployment scenario while remaining simple enough to isolate the effects of orchard geometry and canopy structure on the received signal.

\section{Methodology}

This section presents the methodological framework adopted in this study. It begins by presenting the propagation behavior and anisotropy observed in olive orchards, followed by a detailed description of the geometric environment in the experimental area. %Finally, the simulation design performed in CupCarbon \cite{cupcarbon-website}, is introduced as a preliminary stage prior to real test measurements.

\subsection{Propagation Context and Anisotropy in Olive Orchards}

Orchards, including olive groves, often follow repetitive planting patterns with pruned canopies and fixed row–tree spacing. Depending on the cultivation system, trees can be widely spaced as in traditional groves or tightly planted in rows, as in modern high-density and super-intensive orchards. These variations directly influence the propagation paths and introduce a strong directional dependence in signal attenuation.

\begin{figure}[t]
  \centering
  \includegraphics[width=\columnwidth]{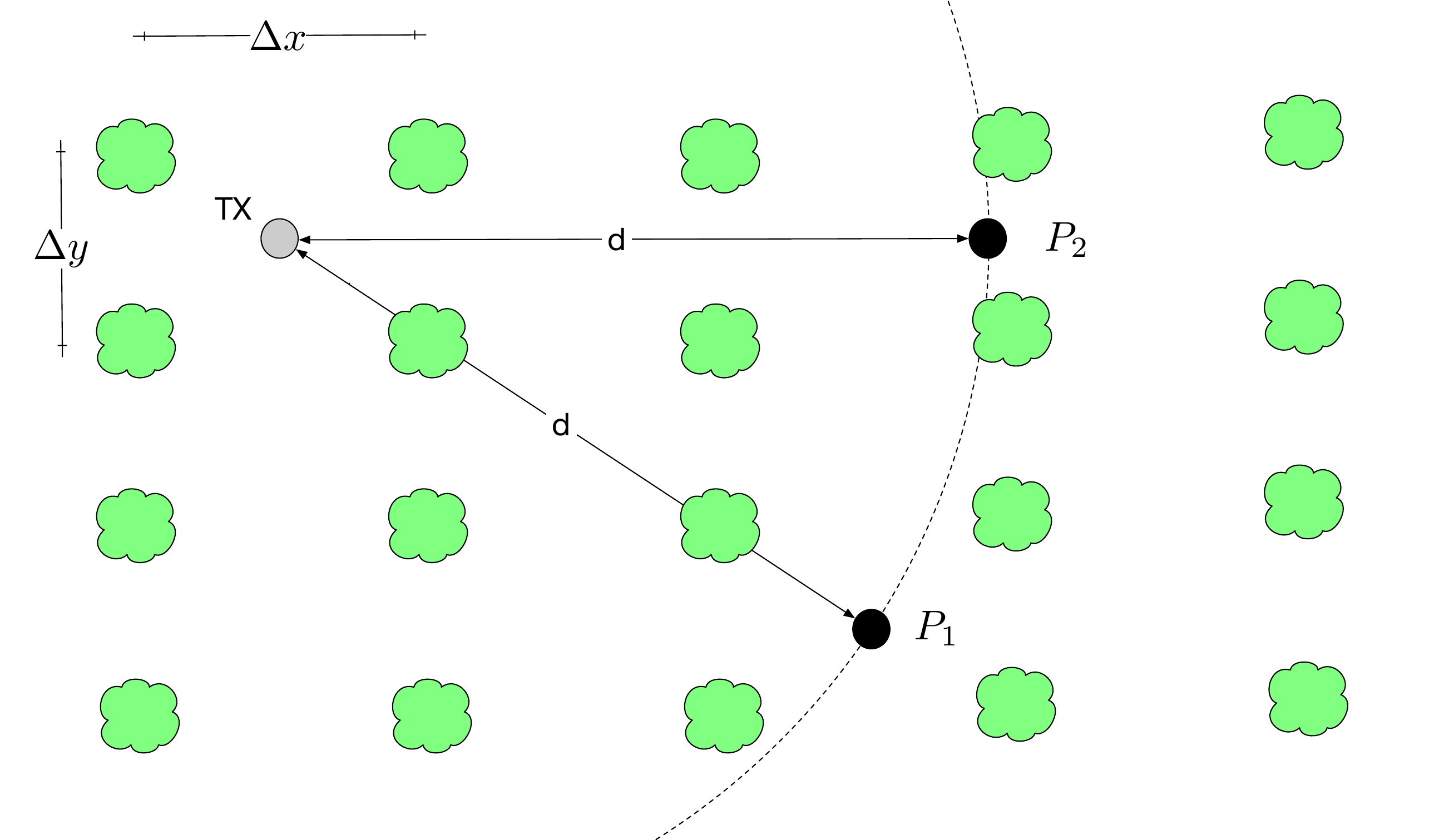}
  \caption{Propagation conditions in two different positions at the same distance from the transmitter when using an omnidirectional propagation model with d as single parameter.}
  \label{fig:propagation_pm_map}
\end{figure}

In conventional isotropic models such as FSPL, ITU-R P.833, or multi-wall, the path loss is assumed to depend only on the Euclidean distance between the transmitter and receiver. However, in a row-based plantation, the radio wave interacts differently when it travels along the rows versus across them, due to the distinct number of tree intersections.
Fig.~\ref{fig:propagation_pm_map} shows this effect: although the receivers $P_{1}$ and $P_{2}$ are located at the same distance $d$ from the transmitter, the signal reaching $P_{1}$ experiences higher attenuation because it crosses several trees, while $P_{2}$ is located on an unobstructed middle corridor. As a result, the loss function $PL(d)$ becomes direction-dependent, revealing the direction-dependent nature of radio propagation in orchards.

%despite distance is the same, $PL(d)_{P_1}>PL(d)_{P_2}$ as the propagation is NLoS in $P_1$ and LoS in $P_2$.

To generalize this behavior, Table.~\ref{tab:olive-systems} summarizes common olive cultivation systems by typical density, row × tree spacing, and canopy arrangement. Denser layouts typical of high-density (HD) and super-high-density (SHD) hedgerow systems form foliage walls that increase cross-row attenuation and reduce the isotropy assumed in classical propagation models.
As the inter-row distance decreases and canopy continuity increases, signal paths are more likely to cross foliage layers rather than travel through open corridors. In traditional construction layouts, trees are separated by open gaps that allow relatively direct paths between transmitter and receiver. In contrast, dense hedgerow systems create continuous foliage corridors where signals are repeatedly attenuated by tree layers. As a result, even when two receiver points are at similar or shorter distances, the one located across multiple rows may experience a stronger loss than a farther point aligned along the rows.

\begin{comment}
\begin{figure*}[h]
  \centering
  \subfloat[Traditional plantation density\label{fig:trad-density}]{
    \includegraphics[height=7cm]{Figures/distances2.pdf}
  } \hfill
  \subfloat[High-density plantation\label{fig:high-density}]{
    \includegraphics[height=7cm]
    {Figures/rows-cols-field-high-density}
  }
  \caption[Traditional plantation density (a); high-density plantation pattern (b). Propagation conditions in two different receiver positions $P_1,P_2$ with our proposed model; $d_1<d_2$ but $A(d_1)>A(d_2)$ because of canopy interception depends on row/cols and not on the Euclidean distance. Unlike literature models, our model can caught this inconsistency because we use two parameters for the attenuation function.]{\protect\colorbox{yellow!35}{\protect\parbox{\dimexpr\linewidth-20\fboxsep\relax}{Traditional plantation density (a); high-density plantation pattern (b). Propagation conditions in two different receiver positions $P_1,P_2$ with our proposed model; $d_1<d_2$ but $A(d_1)>A(d_2)$ because of canopy interception depends on row/cols and not on the Euclidean distance. Unlike literature models, our model can caught this inconsistency because we use two parameters for the attenuation function.}}}
\label{fig:plantation-density}
\end{figure*}
\end{comment}

\begin{figure*}[h]
  \centering
  \subfloat[Traditional plantation density\label{fig:trad-density}]{
    \includegraphics[height=6cm]{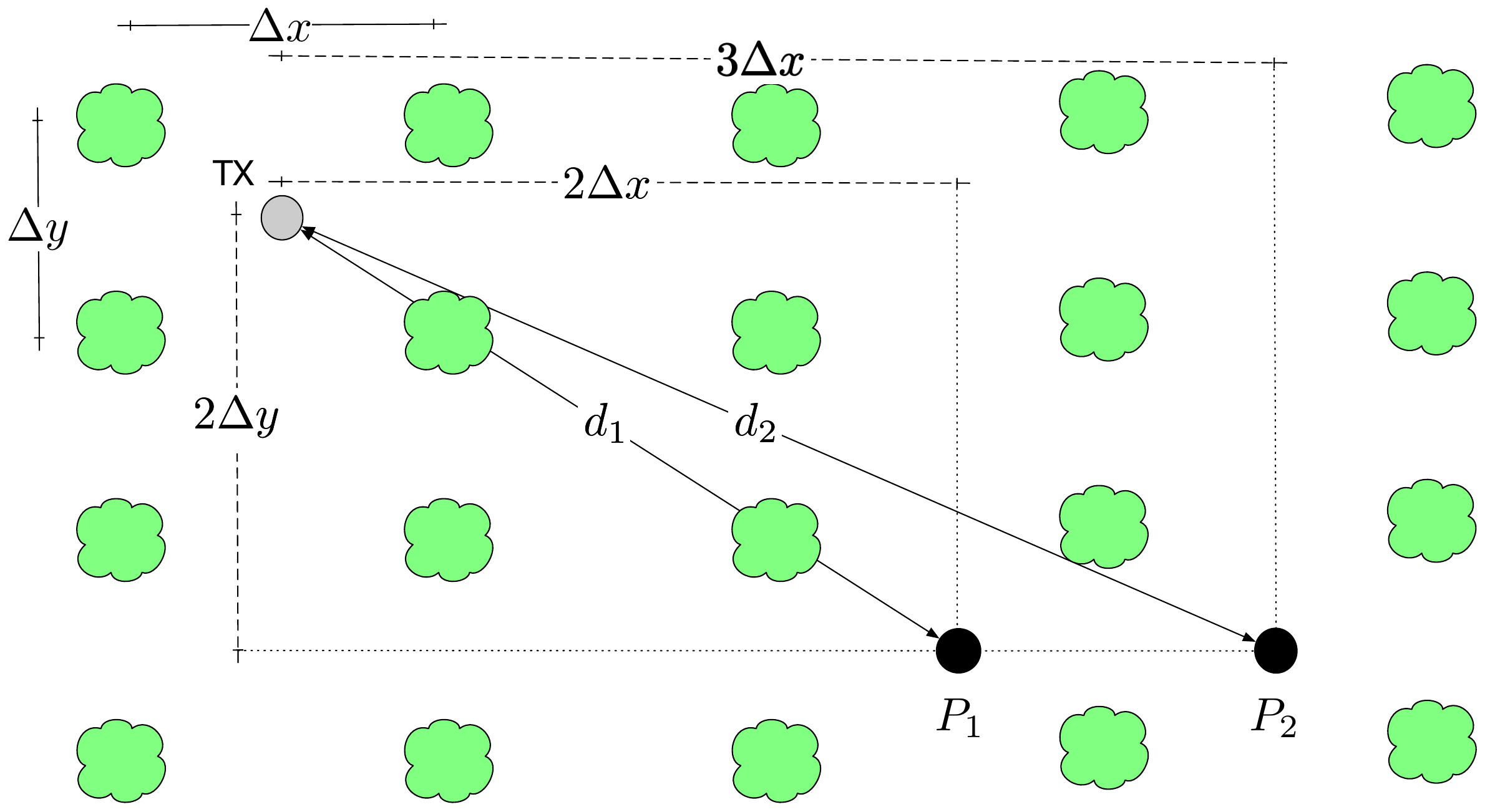}
  } \hfill
  \subfloat[High-density plantation\label{fig:high-density}]{
    \includegraphics[height=6cm]
    {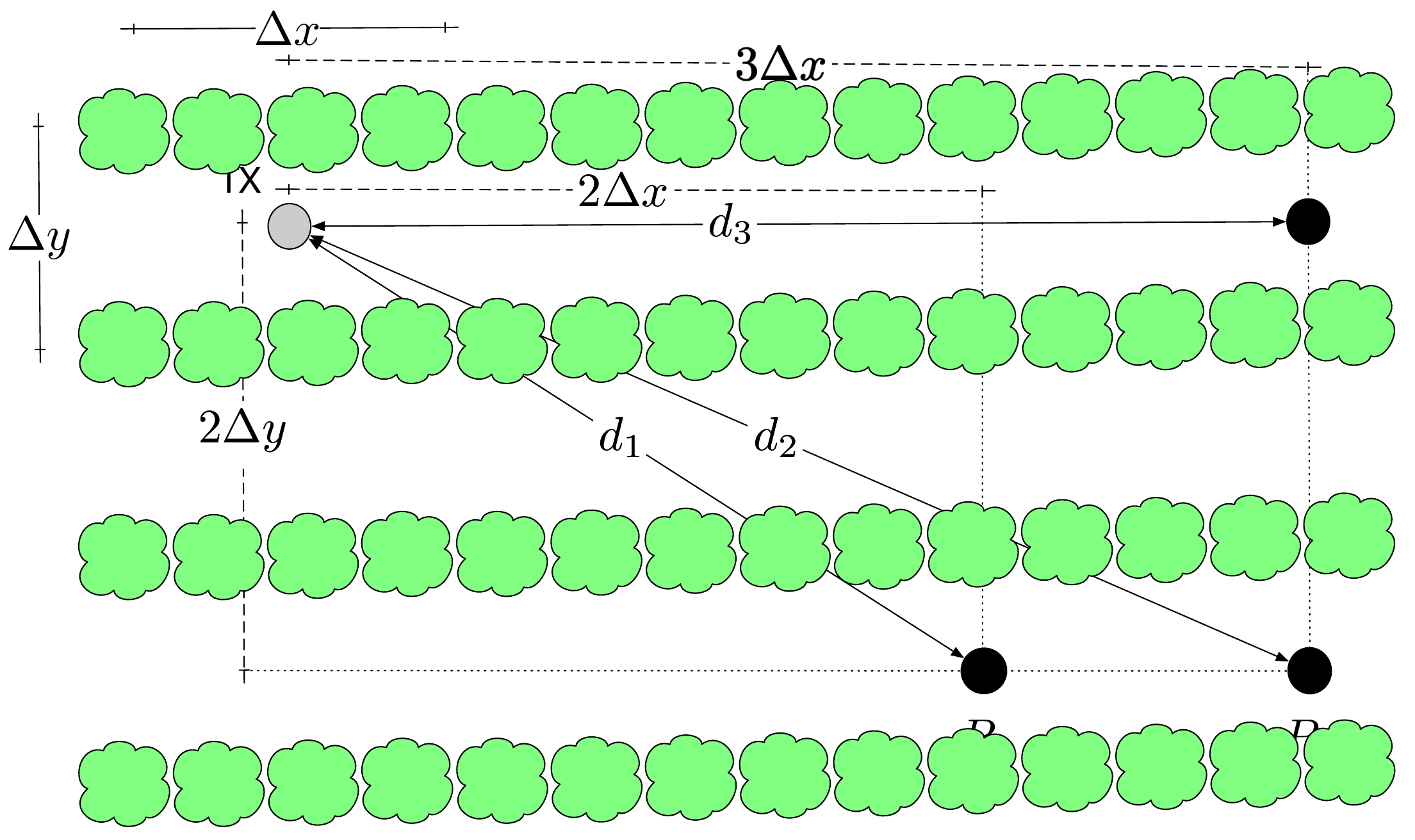}
  }
  \caption[Traditional and high-density plantation patterns illustrating direction-dependent canopy interception.]{
Traditional plantation density (a); high-density plantation pattern (b).
Propagation conditions for two receiver positions $P_1$ and $P_2$ with the proposed model.
Although $d_1<d_2$, the attenuation may satisfy $A(d_1)>A(d_2)$ when the shorter link crosses more canopy layers.
This effect is captured through the canopy-thickness term of the proposed model.
}
\label{fig:plantation-density}
\end{figure*}

\begin{figure}[htbp]
    \centering

    \includegraphics[width=0.78\columnwidth]
    {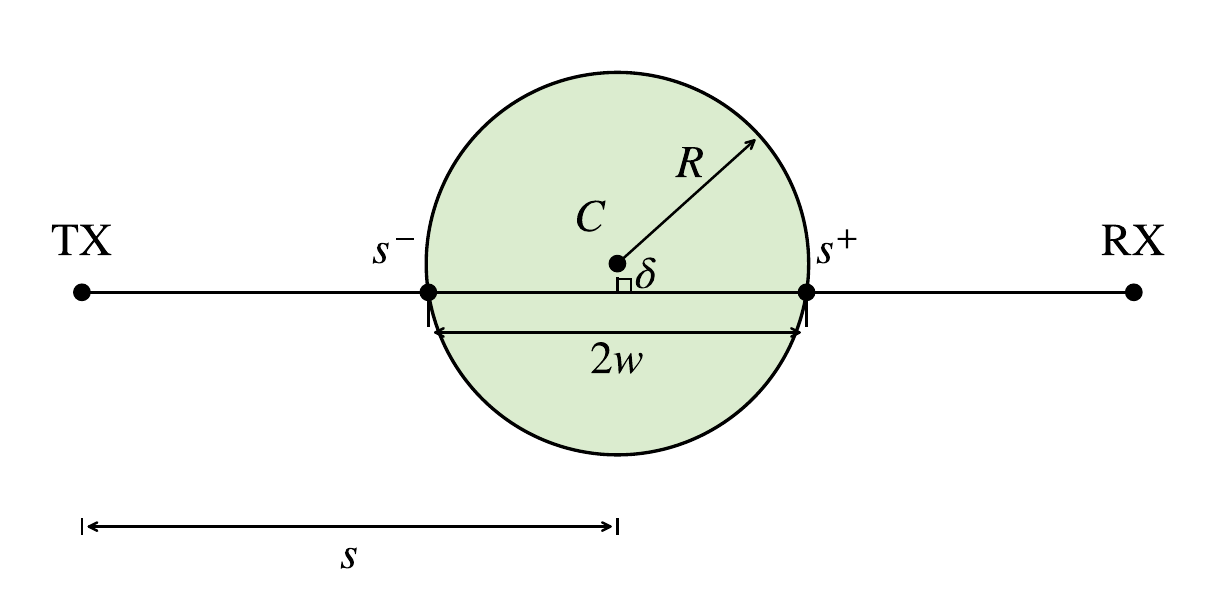}

    \vspace{-1mm}
    {\footnotesize (a) Near-center crossing.}

    \vspace{1mm}

    \includegraphics[width=0.78\columnwidth]
    {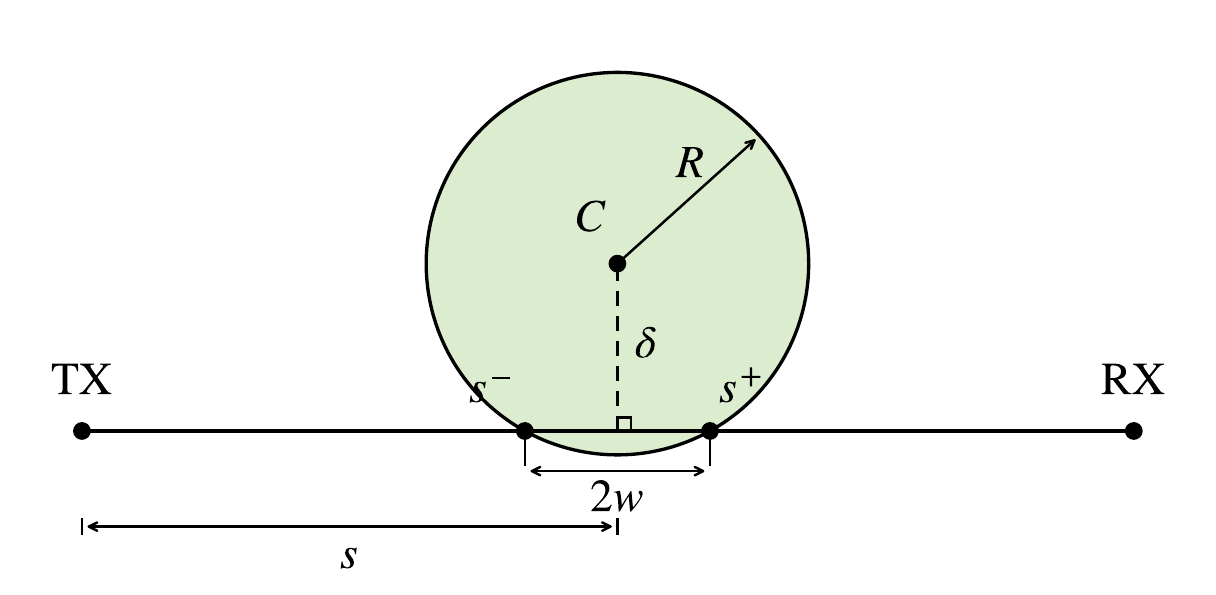}

    \vspace{-1mm}
    {\footnotesize (b) Edge-grazing crossing.}

    \caption{Geometric illustration of the canopy--ray intersection for different crossing conditions.}
    \label{fig:canopy_intersection}
\end{figure}

Figure \ref{fig:plantation-density} shows this behavior: on the left, the traditional configuration shows wider spacing and fewer obstructions, while on the right, the high-density pattern forms compact rows that enhance cross-row attenuation. This contrast emphasizes that propagation in structured orchards cannot be described by distance alone, but depends on the directional components of the field layout.

\begin{figure*}[t]
    \centering
    \begin{minipage}{0.32\linewidth}\centering
        \includegraphics[width=\linewidth]{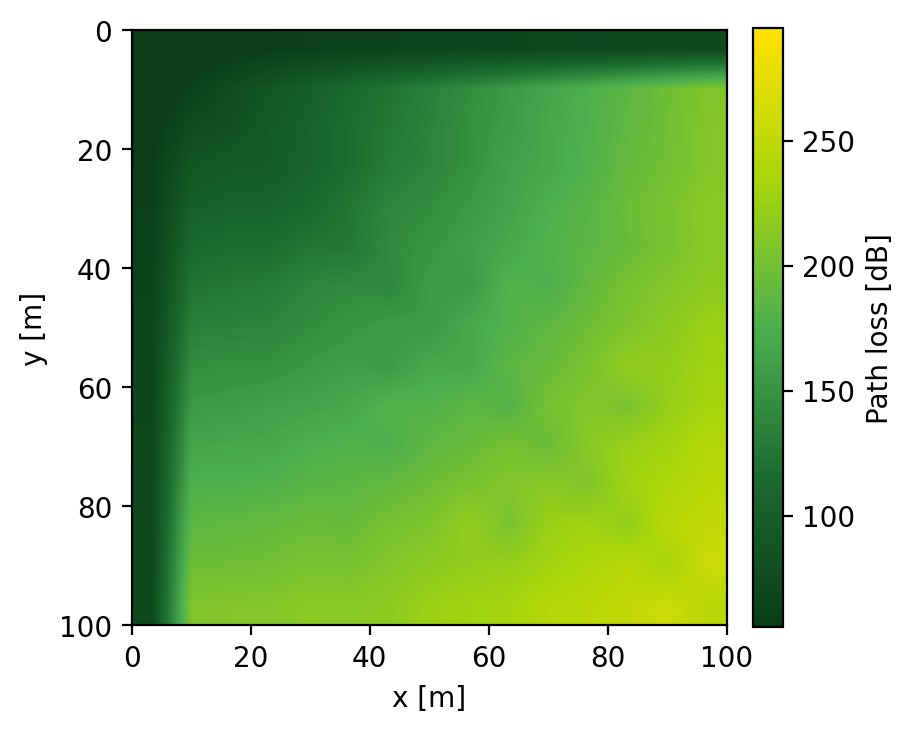}\\
        {\small (a) Traditional}
    \end{minipage}\hfill
    \begin{minipage}{0.32\linewidth}\centering
        \includegraphics[width=\linewidth]{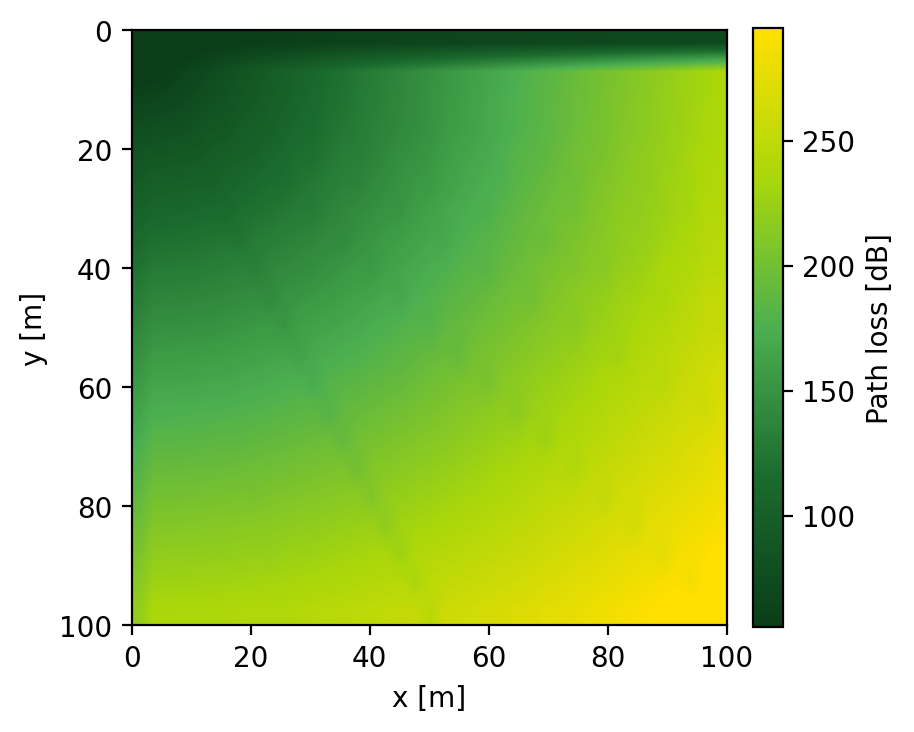}\\
        {\small (b) High-density}
    \end{minipage}\hfill
    \begin{minipage}{0.32\linewidth}\centering
        \includegraphics[width=\linewidth]{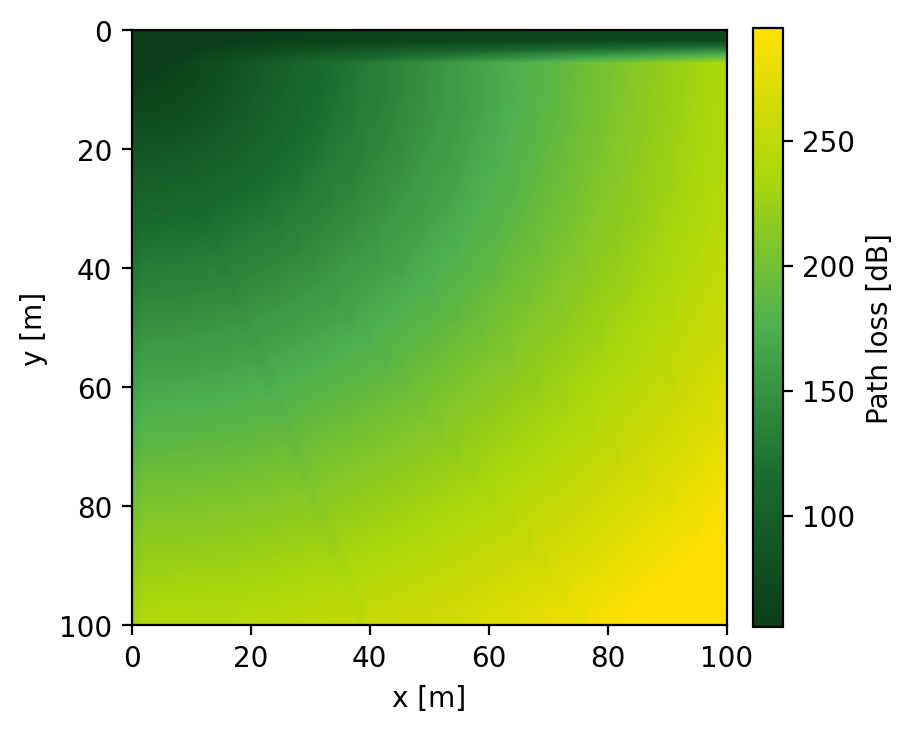}\\
        {\small (c) Super-high-density}
    \end{minipage}
    \caption{Path loss predicted by the model for the three orchard
    classes of a traditional grove (a),
    a high-density planting (b), and a super-high-density hedgerow (c).
    The three maps use the same parameters, field size, and transmitter
    at the top-left corner; only the row and tree spacing differ, and
    the canopy width is scaled to the row spacing so the crowns stay
    within the corridors. Moving from (a) to (c), the rows close up and
    more links are forced to cross foliage, so the loss builds up faster
    away from the transmitter while the low-loss paths survive mainly
    along the open corridors.}
    \label{fig:density_sweep}
\end{figure*}

\subsection{The Proposed Radio Propagation Model}
\label{sec:proposed_model}

In this subsection, we formalize the proposed propagation model as follows:
\begin{equation}
\label{pmodel}
\begin{aligned}
PL(d,L_{\mathrm{tot}})
&=
PL_0
+
10\eta\log_{10}\!\left(\frac{d}{d_0}\right)
+
\alpha L_{\mathrm{tot}},\\
PL_m^{(i)}
&=
PL(d^{(i)},L_{\mathrm{tot}}^{(i)})
+
\epsilon^{(i)} .
\end{aligned}
\end{equation}

In the first relation, the first two terms retain the one-slope
description of the open-field decay, while the third term accounts
for the vegetation actually encountered along the link. Here, $PL(d,L_{\mathrm{tot}})$ denotes the path loss predicted by
the model, $PL_m^{(i)}$ the measured path loss for the $i$th measurement, and $\epsilon^{(i)}$ the corresponding measurement
residual. Here $L_{\mathrm{tot}}$~[m] is the total thickness of the canopy
crossed by the direct ray, and $\alpha$~[dB/m] is the specific foliage
attenuation at the operating frequency. The plantation is described as a
regular rectangular lattice with in-row spacing $\Delta x$ and row spacing
$\Delta y$, so that the canopy center of the tree of indices $(i,j)$ lies
at $(i\,\Delta x,\, j\,\Delta y)$, and every canopy is projected onto the
horizontal plane as a disk of radius $R$, taken to be identical for all trees and
known from the field survey. Expressing the vegetation term as a length
rather than as a count of obstacles removes the quantization that an
integer number of intercepted trees necessarily introduces: a ray grazing
the edge of a crown and a ray passing through its center are no longer
charged the same loss.

Let $\mathrm{TX}=(x_0,y_0)$ and $\mathrm{RX}=(x_1,y_1)$ denote the
transmitter and receiver positions. The direct path between the two devices is the segment
$\mathbf{q}(\xi)=\mathrm{TX}+\xi(\mathrm{RX}-\mathrm{TX})$,
with $\xi\in[0,1]$, whose length is the Euclidean separation given by:
\begin{equation}
d = \lVert \mathrm{RX}-\mathrm{TX} \rVert
  = \sqrt{(x_1-x_0)^2 + (y_1-y_0)^2},
\label{eq:grid_distance}
\end{equation}
where the endpoints are not restricted to lattice nodes: because the
vegetation term is evaluated on the actual straight ray, the same ray has
to set the distance term, and either end may lie anywhere in the plane,
including at the center of a grid cell defined by four neighboring tree positions, or along a corridor between two rows.

Whether a given canopy is crossed follows from two quantities measured
along that ray. Denoting by $\hat{u}=(\mathrm{RX}-\mathrm{TX})/d$ the unit
vector of the path, a canopy centered at $C$ projects onto the path at the
abscissa $s=(C-\mathrm{TX})\cdot\hat{u}$, measured in meters from the
transmitter, and lies at a perpendicular distance $\delta$ from it, with
$\delta^{2}=\lVert C-\mathrm{TX}\rVert^{2}-s^{2}$. The ray misses the
canopy when $\delta \ge R$, and intersects it otherwise. In the latter case, the
intersection is a chord centered on $s$, whose half-length and endpoints are
\begin{equation}
w = \sqrt{R^{2}-\delta^{2}}, \qquad
s^{-} = s-w, \qquad
s^{+} = s+w .
\label{eq:chord}
\end{equation}
Figure~\ref{fig:canopy_intersection}
illustrates the canopy--ray intersection
geometry used to compute the chord length
contributing to $L_{\mathrm{tot}}$. The interval is then clipped to the segment itself, as $\max(s^{-},0)$ and
$\min(s^{+},d)$, so that only the portion lying between the transmitter and
receiver is retained, and it contributes only if it remains non-empty. A
canopy met near its center gives $\delta \to 0$ and a chord approaching the
full diameter $2R$, whereas one met near its edge gives $\delta \to R$ and
a vanishing chord, which is precisely the gradation that a count of
obstacles cannot express.

Summing the chords of all intercepted canopies would charge twice the
foliage shared by crowns that overlap, a situation arising whenever
$\Delta x < 2R$ or $\Delta y < 2R$. The intervals $[s_k^{-},s_k^{+}]$ are
therefore sorted by $s_k^{-}$ and merged whenever
$s_{k+1}^{-}\le s_k^{+}$, and the traversed thickness is taken as the
length of their union $\mathcal{U}$:
\begin{equation}
L_{\mathrm{tot}} = \sum_{[a,b]\,\in\,\mathcal{U}} (b-a) \;\le\; d .
\label{eq:ltot}
\end{equation}
The bound $L_{\mathrm{tot}}\le d$ is a convenient consistency check, since
the foliage traversed can never exceed the length of the link. The merging
step is not merely formal in the orchard under study: with $R = 4.16$~m the
crown diameter is $8.32$~m against a spacing of $7.12$~m in both
directions, so adjacent crowns do intersect, and the union is what keeps
the model from overestimating the foliage on cross-row links.

Because $L_{\mathrm{tot}}$ follows from the exact intersection of the ray
with the lattice of crowns, the directional behavior of the orchard is
reproduced without any explicit angular term. A link traveling along a
corridor stays at a comparatively large perpendicular distance from the
crown centers on either side, so the chords it cuts are short and
$L_{\mathrm{tot}}$ grows slowly with $d$; a link of the same length running
across the rows passes close to one center after another, each
intersection contributing a chord near its maximum length, and the
thickness accumulates far more quickly. Two links with the same separation
but different orientations are thus assigned different losses, which is the
direction-dependent effect that a single scalar distance cannot describe.

The proposed model, therefore, describes the path loss as a two-dimensional
function of the transmitter--receiver geometry over the orchard. Rather
than relying on a single scalar distance, it depends on the layout
$(\Delta x,\Delta y)$, on the canopy radius $R$, and on where the two
endpoints fall within the lattice, so that links of equal length can be
distinguished by their orientation with respect to the rows. Only three
scalar parameters remain to be determined, and the model is linear in all
of them once the geometry is fixed, which keeps it simple enough for
practical use.

Among these, $PL_0$ is the reference path loss at $d_0 = 1$~m, obtained
directly from the received power measured at the reference distance:
\begin{equation}
PL_0 = P_{\mathrm{tx}}^{\mathrm{eff}} - P_r(d_0),
\label{eq:pl0}
\end{equation}
with $P_{\mathrm{tx}}^{\mathrm{eff}}$ the effective transmit power that
absorbs the constant antenna and front-end offsets, yielding
$PL_0 = 28.78$~dB.

Following the one-slope derivation of the multi-wall model~\cite{Anzum2022}, the path-loss exponent
$\eta$ is estimated from the LoS measurement
campaign using the one-slope relation
\begin{equation}
\eta =
\frac{P_r(d_0)-P_r(d)}
{10\log_{10}(d/d_0)} .
\label{eq:eta}
\end{equation}

The value $\eta=2.17$ was obtained from the LoS
measurement campaign with $d_0=1$~m, based on multiple
measurements. Thus, $PL_0$ is obtained from a single
reference reading, whereas $\eta$ is estimated from the
LoS measurement campaign.

For each measurement $i$, the vegetation-related excess loss is then
isolated as the residual left once the reference and distance terms are
removed from the observed path loss as follows:
\begin{equation}
r^{(i)} = PL_m^{(i)} - PL_0 - 10\,\eta \log_{10}\!\left(\frac{d^{(i)}}{d_0}\right),
\label{eq:residual}
\end{equation}
so that, according to~\eqref{pmodel},
$r^{(i)} = \alpha\, L_{\mathrm{tot}}^{(i)} + \epsilon^{(i)}$, with
$\epsilon^{(i)}$ the measurement residual. Since $PL_0$ and $\eta$ are
known and $L_{\mathrm{tot}}^{(i)}$ is fixed by the geometry
of~\eqref{eq:chord} and~\eqref{eq:ltot}, $\alpha$ is the only unknown, and
the problem reduces to a single-parameter regression through the origin.
Minimizing $\sum_i \big(r^{(i)} - \alpha L_{\mathrm{tot}}^{(i)}\big)^2$ and
setting the derivative with respect to $\alpha$ to zero gives the
closed-form least-squares estimate
\begin{equation}
\hat{\alpha} = \frac{\sum_{i} L_{\mathrm{tot}}^{(i)}\, r^{(i)}}
{\sum_{i} \left(L_{\mathrm{tot}}^{(i)}\right)^{2}},
\label{alpha_ls}
\end{equation}
which, over the measurement campaign of this study, yields
$\hat{\alpha} = 1.88$~dB/m. Measurements corresponding to larger vegetation thicknesses exert greater leverage on the estimate of $\alpha$, whereas links with little vegetation contribute comparatively less information.

To illustrate how the proposed model behaves across different planting
systems, Fig.~\ref{fig:density_sweep} reports the predicted path loss over
three orchard layouts of identical size ($100 \times 100$~m), one for each
of the classes listed in Table~\ref{tab:olive-systems}: a traditional
grove, a high-density planting, and a super-high-density hedgerow. The
three maps were generated with the same model parameters and the same
transmitter position, so the only differences between them are the spacing
$(\Delta x,\Delta y)$ and the canopy radius that the corresponding layout
allows. In the traditional orchard, the wide inter-row spacing leaves open
propagation corridors in several directions and the predicted loss remains
moderate over most of the field. As the rows close up, a growing fraction
of the links is forced through foliage, the accumulated thickness rises
faster with distance, and the low-loss paths survive mainly along the
corridors aligned with the rows. The comparison shows that, for the same
area and the same link distances, the path loss is governed not only by
distance but also by the orientation of the link relative to the orchard
rows, which is the direction-dependent effect that the proposed model is
designed to capture.

\subsection{Geometry and Description of the Study Area}

\begin{figure}[t]
    \centering
    \includegraphics[width=0.9\linewidth]{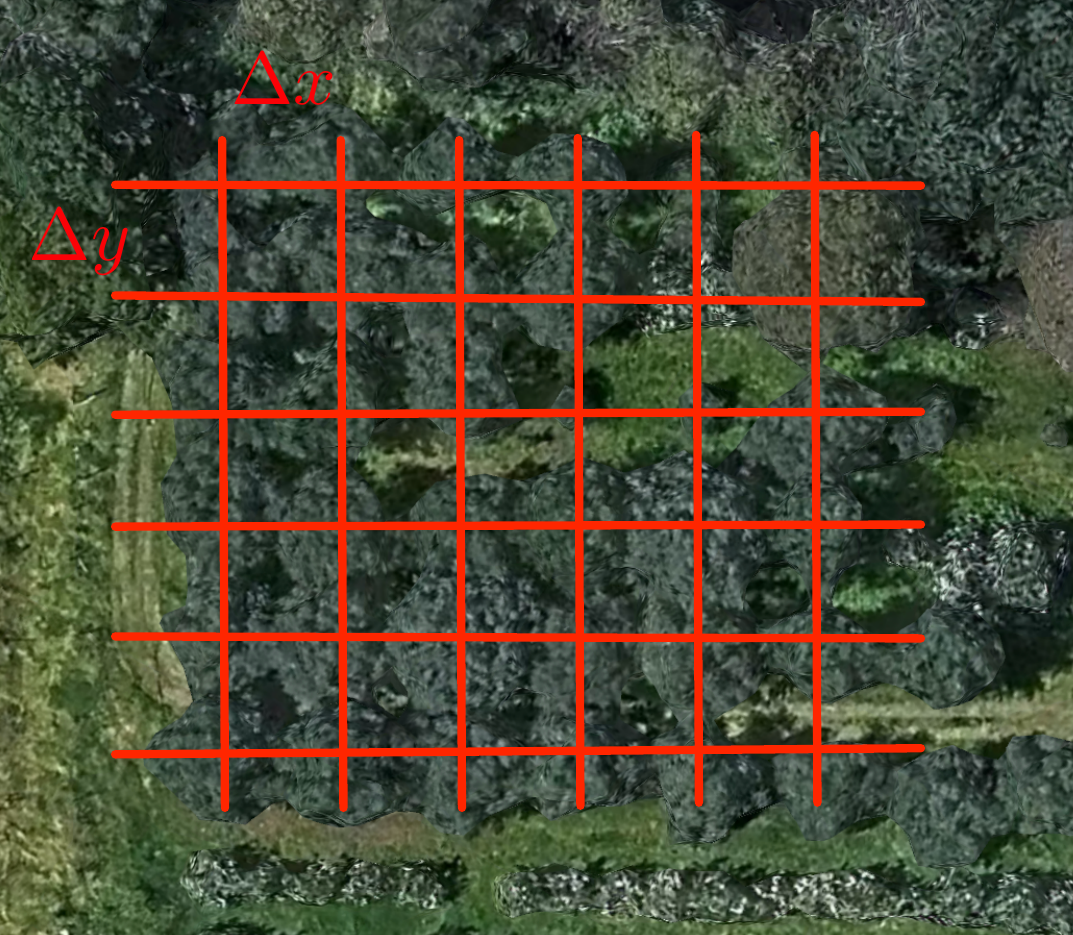}
    \caption{ Satellite view showing the plantation pattern of the experimental field 
    %site in Palermo, 38,10664° N, 13,34976° E. 
    Imagery ©2025 Maxar Technologies, Airbus. Map data ©2025 Apple. Source: Apple Maps\cite{applemaps}.
    }
    \label{fig:battery_box31}
\end{figure}

In the experimental setup, we focus on an olive field located in Palermo, Sicily, Italy (cf. Fig. \ref{fig:battery_box31}). The field is located at coordinates 38,10664° N, 13,34976° E, with a typical Mediterranean farming area with a square pattern of olive trees, some uneven terrain, and moderate vegetation density. The dimensions of the plantation are 43 meters by 38 meters. These dimensions were obtained directly from satellite imagery and mapping tools to accurately represent the actual layout of the field used in this study. This field corresponds to a traditional plantation layout, characterized by moderate spacing and discrete trees, consistent with the parameters of conventional Mediterranean orchards \cite{Connor2014}.

The field is largely flat, with only minor variations in terrain elevation, ensuring near line-of-sight visibility between adjacent rows. The trees form a uniform geometric grid roughly aligned north-south. 
%and the surrounding area is free of major obstructions such as buildings. 
This configuration allows the site to serve as a controlled environment for analyzing propagation anisotropy owing to canopy interception rather than topographic effects.

\section{Experimental Setup}

\begin{figure}[t]
    \centering
    \includegraphics[width=0.95\columnwidth]{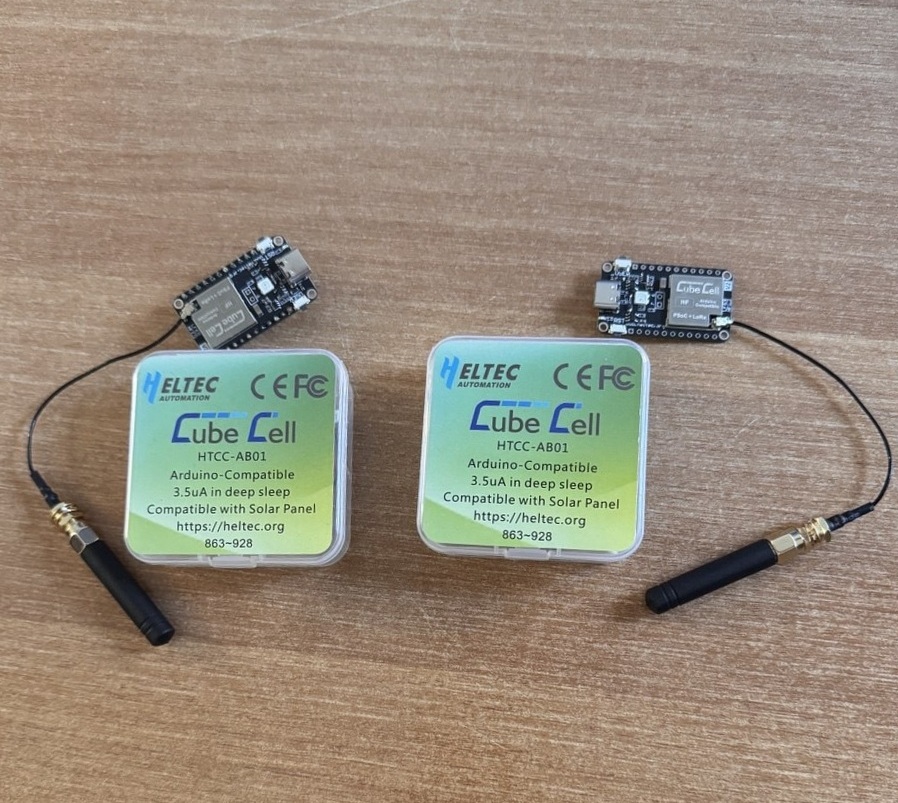}
    \caption{Heltec CubeCell HTCC-AB01 LoRa modules used in our measurements \cite{HeltecHTCCAB01V2}.}
    \label{fig:cubecell_modules}
\end{figure}

This section describes the field measurements used to characterize LoRa propagation in an olive orchard in Palermo (Sicily, Italy). The quantitative model comparison focuses on the reference mid-corridor configuration, with both the transmitter and receiver antennas mounted at 1.2~m above ground. Each node was a Heltec CubeCell HTCC-AB01 (SX1262-class) configured for operation in the European 863--870~MHz band (EU-868), as shown in Fig.~\ref{fig:cubecell_modules}.

% \begin{figure}[H]
%     \centering
%     \includegraphics[width=\linewidth]{Figures/palermo farm.jpg}
%     \caption{ Plantation pattern and view of the experimental field site in Palermo, Sicily, Italy.}
%     \label{fig:battery_box31}
% \end{figure}

%\begin{figure}[H]
 %   \centering
  %  \includegraphics[width=\linewidth]{Figures/palermo,freearea.jpg}
  %  \caption{ Satellite view of the experimental open area site in Palermo, Sicily, Italy. Imagery ©2025 Airbus, Maxar Technologies. Map data ©2025 Google \cite{googlemaps}.}
  %  \label{fig:battery_box31}
% \end{figure}

\subsection{Measurement Configuration and Procedure}

%The measurements were carried out on a regular-grid olive orchard. 
\begin{comment}
\noindent
{\setlength{\fboxsep}{6pt}%
\colorbox{yellow!35}{%
\parbox{\dimexpr\linewidth-2\fboxsep-2\fboxrule\relax}{%
All field measurements were conducted in September and October under a single canopy condition of the olive orchard. Since olive trees, Olea europaea L., are an evergreen species. The timing of the campaign is also relevant to the intended application, since it leads into and partly overlaps the main olive-harvesting period in Italy, which typically extends from mid-October through November.
%, with leaves persisting for two or more years in the canopy
and the orchard did not exhibit a true leaf off state comparable to deciduous orchards. Therefore, the present study does not separate attenuation into leaf on and leaf off cases. Instead, vegetation loss is represented through the measured orchard geometry, namely the canopy intersections along the Tx--Rx path and the representative mean canopy radius measured in the field. The proposed model is therefore calibrated for the canopy condition observed during the reported campaign.%
}}}\cite{Connor2014}
\end{comment}

All field measurements were conducted in September and October 2025 under a single canopy condition of a regular-grid olive orchard. Since olive trees (\textit{Olea europaea} L.) are evergreen, the orchard did not exhibit a true leaf-off state comparable to deciduous orchards. The campaign also preceded and partly overlapped the main olive-harvesting period in Italy, which typically extends from mid-October through November. Therefore, the present study does not separate attenuation into leaf-on and leaf-off cases. Instead, vegetation loss is captured through the measured orchard geometry, namely canopy intersections along the Tx--Rx path and the mean canopy radius measured in the field. The proposed model is therefore calibrated for the canopy condition of the reported campaign \cite{Connor2014}.

%Two adjacent rows were selected as the baseline corridor for the first test,  
%the second test added cross-row traverses and dual antenna heights, with measurements taken both close to the trunk (under the canopy) and in the inter-row corridor, in the midpoint between two olive trees. 
In the %third
experiment, the entire orchard was systematically surveyed along mid-row and under-canopy trajectories.  
The orchard exhibits near-uniform spacing along both axes; average inter-row and intra-row spacings are shown in Table~\ref{tab:site-params}.

\begin{table}[h]
\caption{Site parameters of the experimental olive orchard}
\label{tab:site-params}
\centering
\begin{tabular}{l c}
\hline
Location & Palermo, Sicily, Italy \\
Inter-row spacing $s_{\text{row}}$ (m) & 7.12 \\
Intra-row spacing $s_{\text{col}}$ (m) & 7.12 \\
Representative canopy radius $r_c$ (m) & 4.16 \\
\hline
\end{tabular}
\end{table}

\begin{comment}
    
\noindent
{\setlength{\fboxsep}{6pt}%
\colorbox{yellow!35}{%
\parbox{\dimexpr\linewidth-2\fboxsep-2\fboxrule\relax}{%
In the present study, vegetation attenuation is represented through the measured orchard geometry, namely the canopy intersections along the Tx--Rx path and the representative mean canopy radius measured in the field. However, the proposed formulation does not introduce a separate local term for links propagating predominantly inside dense foliage. Accordingly, the model is intended to describe large scale direction dependent attenuation in the structured olive orchard, while the main model comparison is performed for the reference mid corridor configuration.%
}}}
\end{comment}

In the present study, vegetation attenuation is represented through the measured orchard geometry, namely the canopy intersections along the Tx--Rx path and the representative mean canopy radius measured in the field. However, the proposed formulation does not introduce a separate local term for links propagating predominantly inside dense foliage. Accordingly, the model is intended to describe large scale direction dependent attenuation in the structured olive orchard, while the main model comparison is performed for the reference mid corridor configuration.

In our setup, both the transmitter and the receiver used identical omnidirectional monopole antennas of approximately 48\,mm length, operating at 868\,MHz. At 868\,MHz, these antennas are shorter than a half-wave dipole ($\lambda/2 \approx 172$\,mm). However, their realized gains were not independently characterized, so a specific numerical gain cannot be inferred from antenna length alone. Since both link ends used the same antenna type, the antenna-related contribution is treated as a setup-dependent offset in the link budget. For example, a 1\,dB change in gain at each antenna would produce a 2\,dB change in received power, assuming all other conditions remain unchanged. A separate antenna-gain term is therefore omitted in our formulation, and all fixed front-end effects are implicitly absorbed into an effective transmit power defined over the horizontal 2D plane. Because the same antennas were used throughout the campaign, this offset is treated as constant for all measured links and does not affect the comparative evaluation of the considered propagation models. The calibrated model should accordingly be interpreted within the canopy, antenna-height, and hardware conditions of the reported campaign: variations in leaf moisture, pruning state, canopy density, antenna height, or antenna gain may shift the absolute RSSI level and the calibrated parameter values, and a different antenna or front-end configuration would require recalibrating the effective transmit power and canopy-loss terms. Such variations do not, however, change the geometric structure of the proposed formulation, which rests on the Tx--Rx geometry and canopy intersections.

\begin{comment}
    
\noindent
{\setlength{\fboxsep}{6pt}%
\colorbox{yellow!35}{%
\parbox{\dimexpr\linewidth-2\fboxsep-2\fboxrule\relax}{%
RSSI values were recorded through serial monitoring. At each receiver position, 30 samples were collected and averaged to reduce instantaneous fluctuations caused by small-scale fading. The resulting value was then rounded to represent the effective signal level at that point.}}} 
\end{comment}

RSSI values were recorded through serial monitoring. At each receiver position, 30 samples were collected and averaged to reduce instantaneous fluctuations caused by small-scale fading. The resulting value was then rounded to represent the effective signal level at that point.

The configuration summarized in Table~\ref{tab:radio-params} was used for all measurements to ensure stability between measurement campaigns. The Tx/Rx nodes were programmed using a custom Arduino script \cite{rowhani2025anisotropic}, which automated packet transmission and reception under the same LoRa parameters. The script defined a solid transmission interval, recording each received packet timestamp and RSSI value. This approach guaranteed identical behavior of the transmitter and receiver during all field tests.

\begin{table}[htb]
\caption{LoRa radio and sampling setup}
\label{tab:radio-params}
\centering
\begin{tabular}{l c}
\hline
Node model & Heltec CubeCell HTCC-AB01 (SX1262) \\
Operating band & EU 863--870\,MHz (EU-868) \\
Bandwidth & 125\,kHz \\
Coding rate & 4/5 \\
Spreading factor & SF10 \\
Tx power $P_{\mathrm{Tx}}$ & 21\,dBm \\
Antenna & Omnidirectional monopole \\
Packets per waypoint & $\geq 30$ (constant payload, constant interval) \\
Receiver logs & waypoint ID, timestamp, per-packet RSSI \\
\hline
\end{tabular}
\end{table}

\begin{comment}
    
\noindent
{\setlength{\fboxsep}{6pt}%
\colorbox{yellow!35}{%
\parbox{\dimexpr\linewidth-2\fboxsep-2\fboxrule\relax}{%
In all tests, the transmitter was mounted fixed at 1.2~m above ground. Measurements were taken along two paths: the mid-corridor between rows and under-canopy near the trunk. Figure~\ref{fig:los_corridor} illustrates only the reference mid-corridor configuration with both antennas at 1.2~m. After comparing the datasets across height and path, we selected this configuration as the baseline case for the main model comparison reported in this work.%
}}}
\end{comment}

\begin{figure}[t]
    \centering
    \includegraphics[width=\linewidth,  trim=0cm 3cm 0cm 12cm,
  clip]{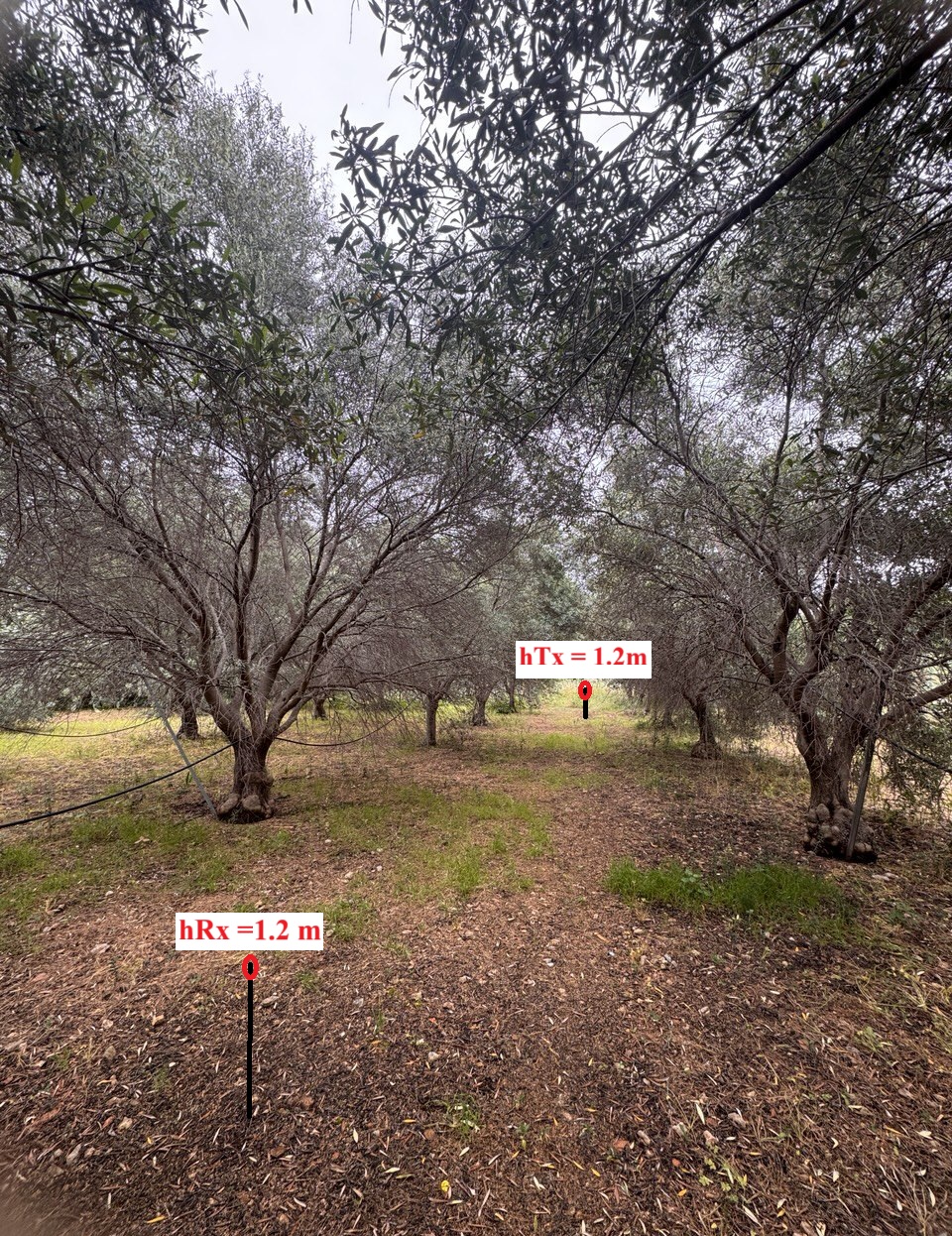}
    \caption{Line-of-sight propagation corridor in the experimental olive orchard, with Tx and Rx antennas at 1.2 m above ground.}
    \label{fig:los_corridor}
\end{figure}

For the reference measurement configuration used in the quantitative comparisons reported in this study, both transmitter and receiver antennas were positioned at 1.2~m above ground, along the mid-corridor between tree rows. Figure~\ref{fig:los_corridor} illustrates this configuration. The RSSI heatmaps and model-error metrics reported in Section~\ref{sec:propagation_models} correspond to this reference configuration. Additional experimental results and datasets are available in the GitHub repository associated with this study \cite{rowhani2025anisotropic}.

\section{Numerical results}
\label{sec:propagation_models}

\begin{figure}[t]
  \centering
  \includegraphics[width=0.35\textwidth]{\detokenize{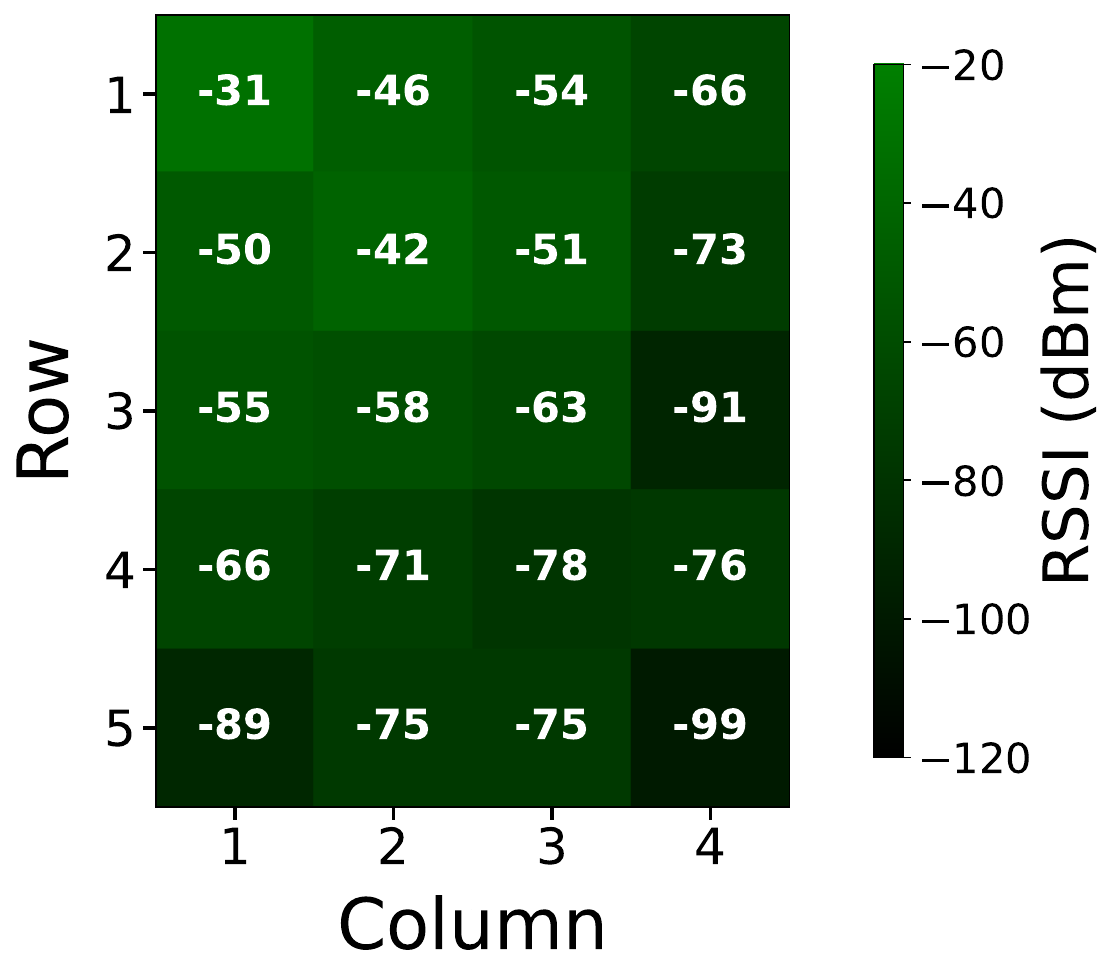}}
\caption[Olive orchard measured RSSI heatmap]{Olive orchard measured RSSI heatmap}
\label{fig:real_rssi}
\end{figure}

To understand how different propagation models reproduce the signal distribution measured in the olive orchard (Fig.~\ref{fig:real_rssi}), we compare their predictions directly against the experimental RSSI data sets. Three representative formulations are considered in detail: the ITU--R vegetation model, a multi-wall obstruction model, and our proposed 2-D direction-dependent model (cf. Section~\ref{releatedworks}). In all cases, the models are evaluated on the same transmitter--receiver geometry used during the field experimental tests, so that their outputs can be contrasted point by point with the measurements.

\begin{comment}
\begin{figure}[t]
  \centering
  \includegraphics[width=0.4\textwidth]{\detokenize{Figures/Heatmaps/the_latest_heatmaps/01_real_rssi.pdf}}
\caption[Olive orchard measured RSSI heatmap]{\protect\colorbox{yellow!35}{\protect\parbox{\dimexpr\linewidth-20\fboxsep\relax}{Olive orchard measured RSSI heatmap}}}
\label{fig:real_rssi}
\end{figure}
\end{comment}

% Two other models introduced earlier in the section \ref{releatedworks}  are deliberately kept outside this comparison. The free-space path loss (FSPL) model is used as an ideal reference to decompose the measured attenuation and to calibrate baseline terms, but it assumes unobstructed line-of-sight and isotropic spreading. In a structured orchard, it cannot be fully compared for induced losses, or directional effects can't be considered as a realistic candidate model. 
%We compare the experimental results (measures collected in the field) with the results of the most significant models as discussed in Section~\ref{releatedworks}: ITU-R and multi-wall, then with our proposed model. 
 
% Also, the excess-loss term \(Z(d)\), obtained by subtracting the FSPL baseline from the measured attenuation, is a diagnostic quantity built from the data themselves. It is useful to isolate vegetation related loss and to guide the design of more refined models, but it does not generate independent RSSI prediction datas. For these reasons, both FSPL and \(Z(d)\) are excluded from the quantitative comparison, which focuses on models that provide stand-alone predictions on the measurement grid.

\begin{figure*}[htbp]
  \centering
  \subfloat[\label{fig:itu-model}]{
    \includegraphics[width=0.34\textwidth]{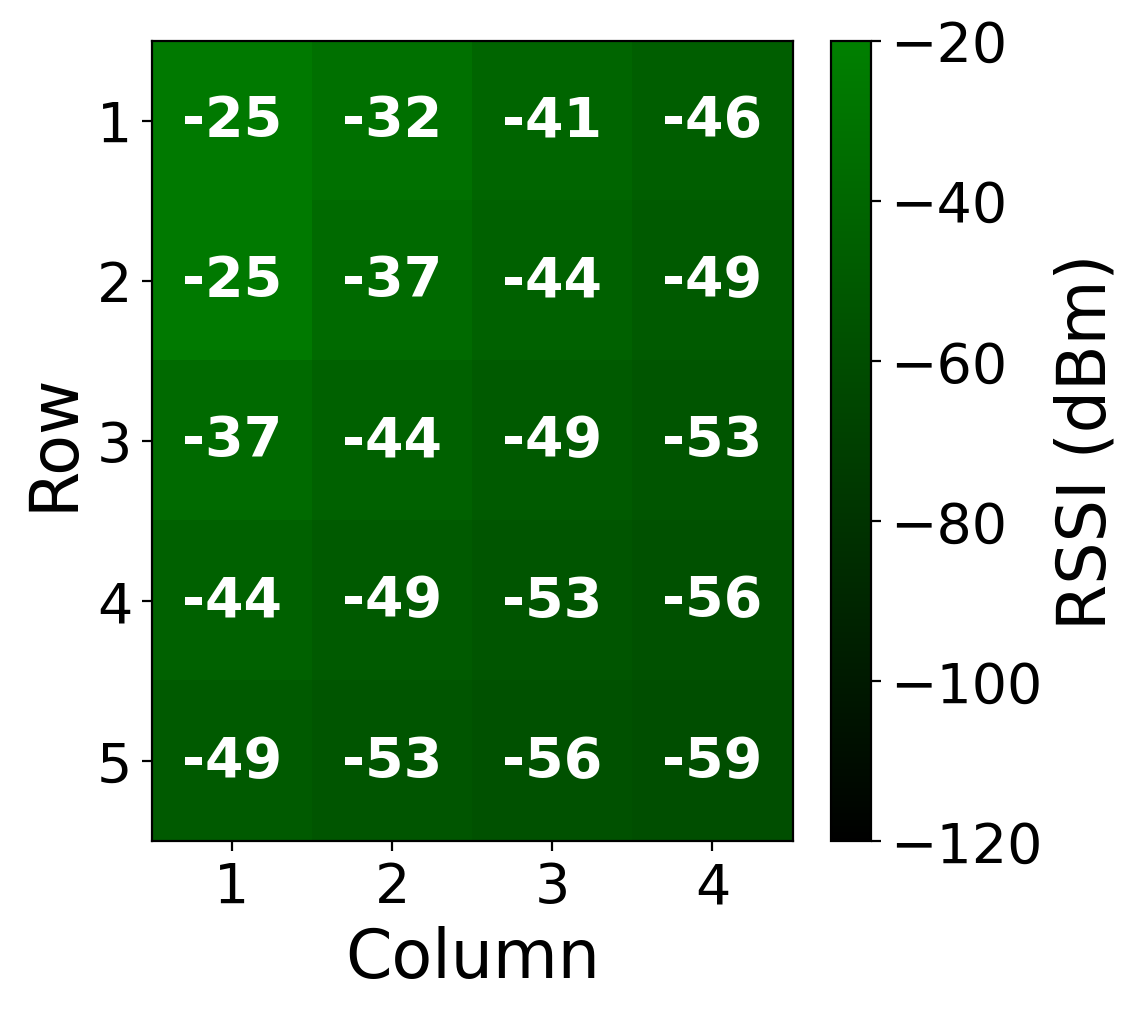}
  }
  \subfloat[\label{fig:itu-error}]{
    \includegraphics[width=0.34\textwidth]{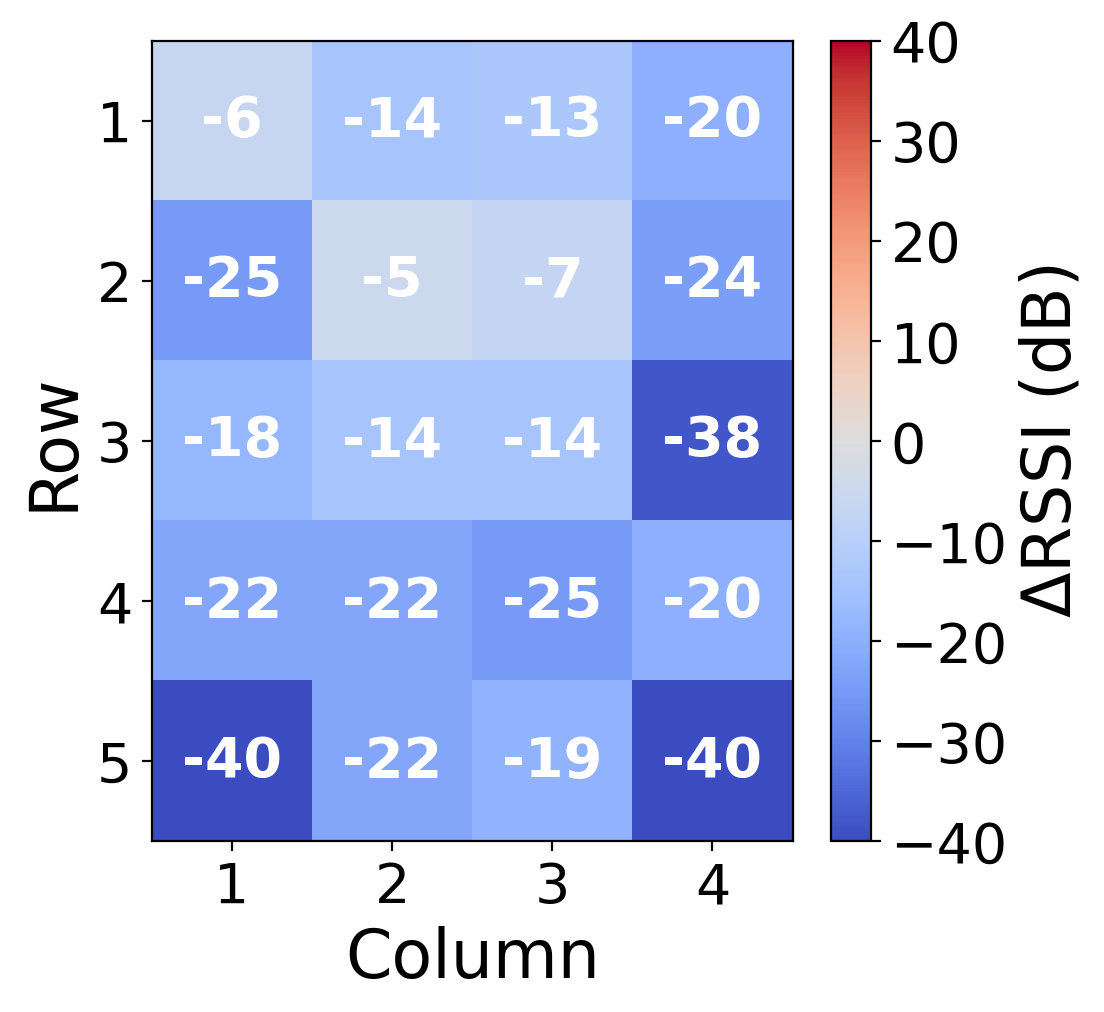}
  }\\
  \subfloat[\label{fig:multiwall-model}]{
    \includegraphics[width=0.34\textwidth]{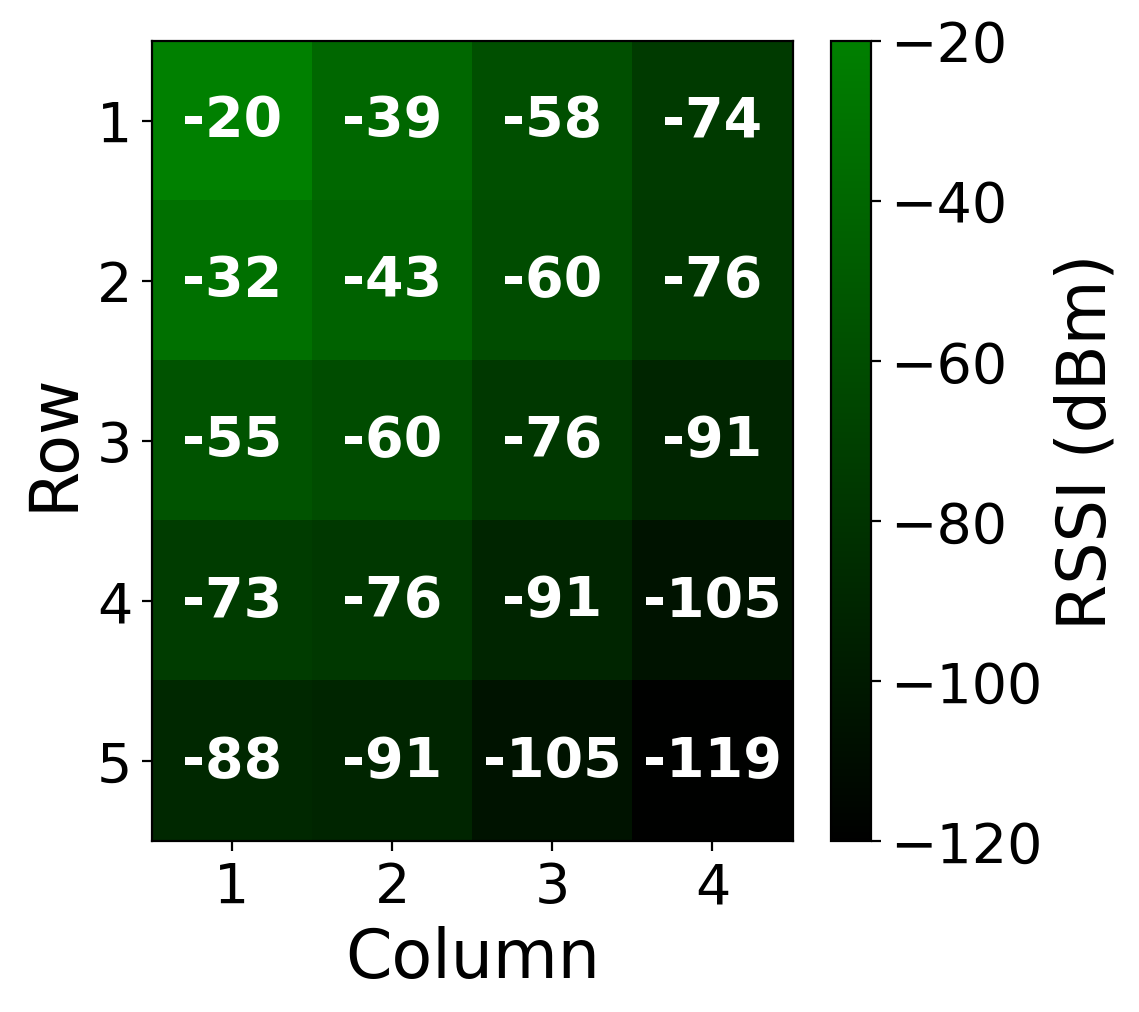}
  }
  \subfloat[\label{fig:multiwall-error}]{
    \includegraphics[width=0.34\textwidth]{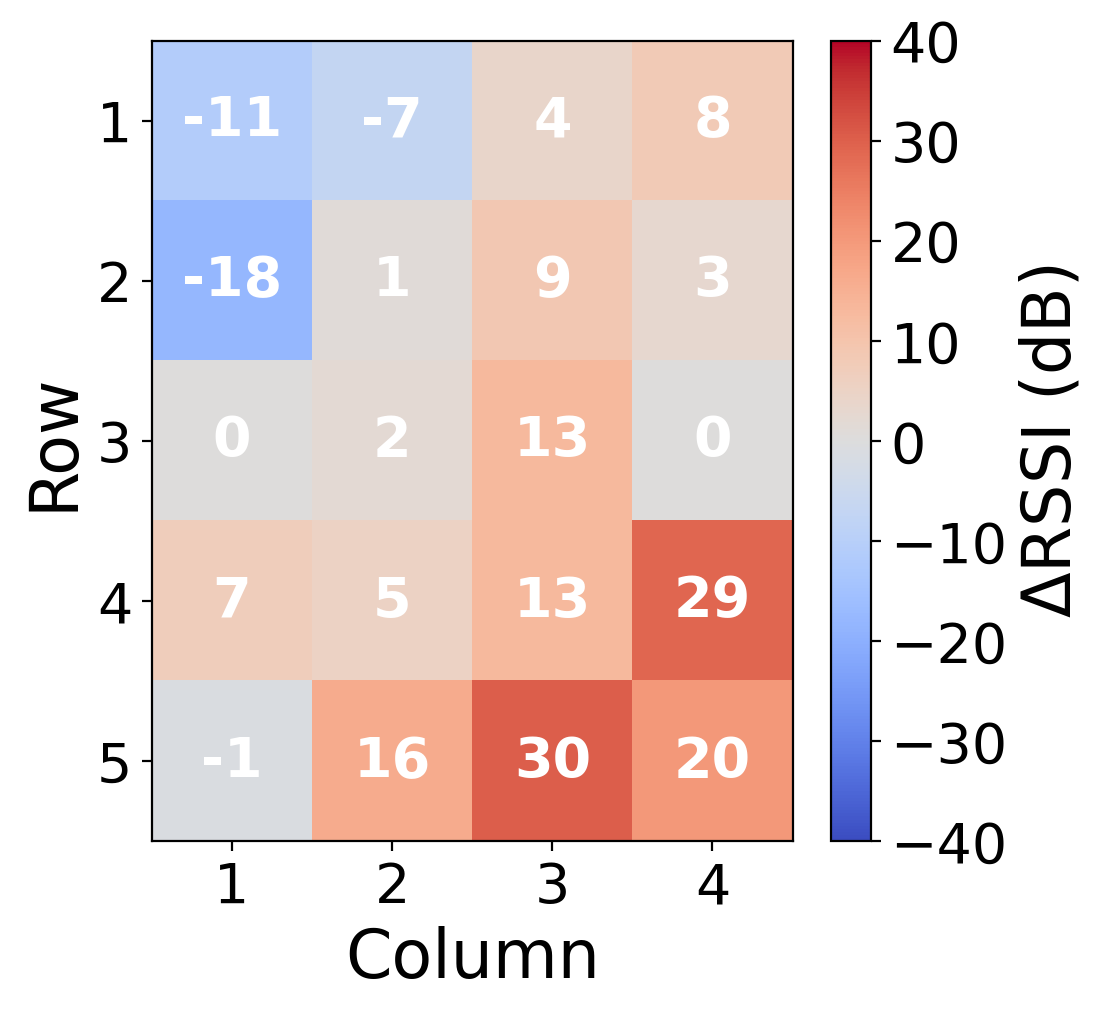}
  }\\
  \subfloat[\label{fig:proposed-model}]{
    \includegraphics[width=0.34\textwidth]{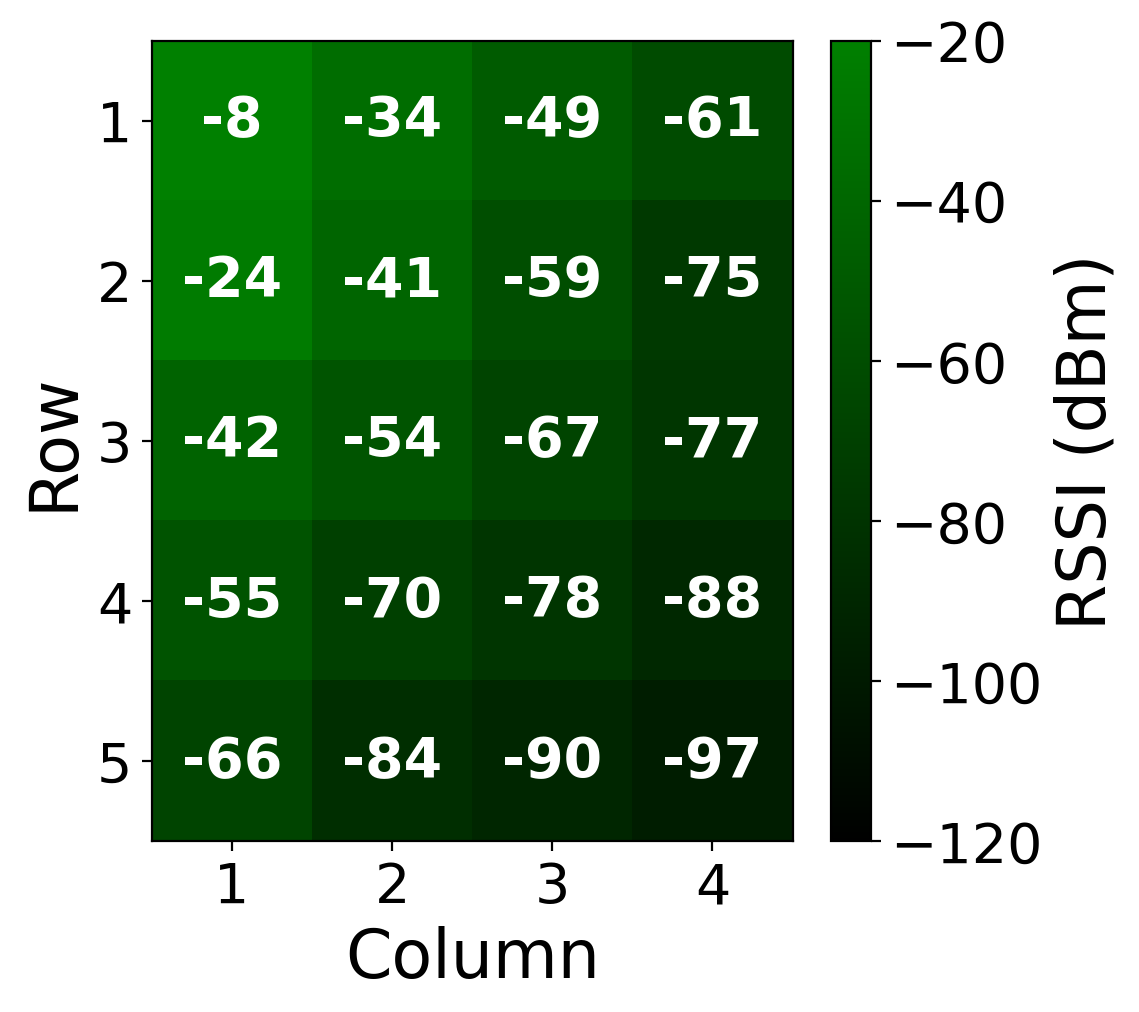}
  }
  \subfloat[\label{fig:proposed-error}]{
    \includegraphics[width=0.34\textwidth]{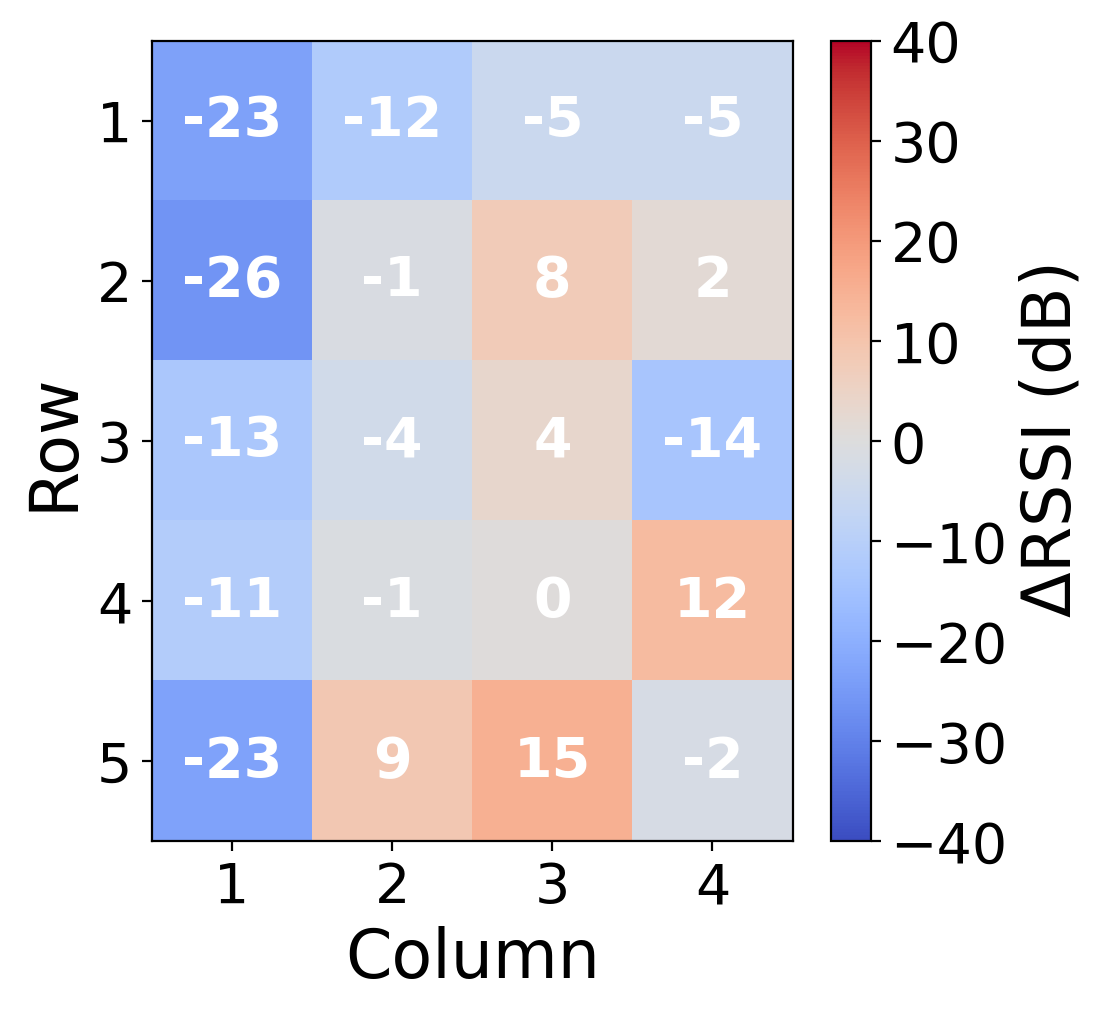}
  }

\caption[Comparison of modeled and measured RSSI for the mid-corridor scenario at 1.2~m height: the ITU-R (a), multi-wall (c), the proposed 2-D direction-dependent model (e) and their respective error heatmaps (b), (d), and (f).]{Comparison of modeled and measured RSSI for the mid-corridor scenario at 1.2~m height: the ITU-R (a), multi-wall (c), the proposed 2-D direction-dependent model (e) and their respective error heatmaps (b), (d), and (f).}
\label{fig:results}
\end{figure*}

Each propagation model was evaluated at the same transmitter--receiver measurement locations, producing its own RSSI heatmap over the grid. This makes it possible to inspect how each model represents the signal distribution across and along the tree rows. Then, the point-wise difference between measured and modeled RSSI is computed to obtain an error heatmap, which highlights where and to what extent each model departs from the observed behavior, as in Fig.~\ref{fig:results}.

%%%%%%%% end template

To complement visual inspection, the mean square error (MSE) and the root mean square error (RMSE) were calculated for each model over the full set of measurement points. These indicators condense the pointwise discrepancies into single figures and provide a compact basis for comparison. The results are summarized in Table~\ref{tab:rmse}.

\begin{comment}
    
{\setlength{\fboxsep}{6pt}%
\colorbox{yellow!35}{%
\parbox{\dimexpr\linewidth-2\fboxsep-2\fboxrule\relax}{%
Table~\ref{tab:rmse} and Fig~\ref{fig:results}
provide a direct comparison between the proposed formulation and the representative baseline models evaluated in this study under the same mid-corridor measurement geometry. In addition to the lower overall error reported in Table~5, Fig.~9 shows that the proposed direction-dependent model follows the measured RSSI distribution over the orchard grid more closely, with smaller and less structured residual regions than the ITU--R vegetation and plantation multi-wall models. 
These results indicate that explicitly accounting for orchard geometry improves the agreement between the modeled and measured signal distribution in the studied olive orchard.%
}}}

\end{comment}

Table~\ref{tab:rmse} and Fig~\ref{fig:results}
provide a direct comparison between the proposed formulation and the representative baseline models evaluated in this study under the same mid-corridor measurement geometry. In addition to the lower overall error reported in Table~5, Fig.~\ref{fig:results} shows that the proposed direction-dependent model follows the measured RSSI distribution over the orchard grid more closely, with smaller and less structured residual regions than the ITU--R vegetation and plantation multi-wall models. 
These results indicate that explicitly accounting for orchard geometry improves the agreement between the modeled and measured signal distribution in the studied olive orchard.

The ITU--R vegetation model exhibits the largest error, with an RMSE of 22.91~dB, which is consistent with its isotropic structure and its limited
ability to reproduce the direction-dependent attenuation pattern observed in Fig.~\ref{fig:results}. Introducing canopy-related losses through discrete obstacle counts improves the comparison: the multi-wall model reduces the RMSE to 13.18~dB, showing that an explicit treatment of foliage intersections is beneficial in this setting. The proposed 2-D direction-dependent formulation further reduces the RMSE to 12.20~dB and
produces a visibly more homogeneous error map. As a result, it provides a closer description of the directional attenuation patterns that characterize
the olive orchard.

\begin{table}[t]
  \centering
  \caption{MSE and RMSE of our model with respect to experimental measurements.}
  \label{tab:rmse}
  \begin{tabular}{lcc}
    \hline
    \textbf{Model} & \textbf{MSE (dB$^{2}$)} & \textbf{RMSE (dB)} \\
    \hline
    ITU--R vegetation          & 524.68 & 22.91 \\
    Multi-wall                 & 173.79 & 13.18 \\
    Proposed 2-D direction-dependent   & 148.73 & 12.20 \\
    \hline
  \end{tabular}
\end{table}

\section{Conclusion}
This study examined radio propagation in a real olive orchard and showed that structured fields play a decisive role in shaping LoRa behaviors. Field measurements conducted to represent communication between a tractor-mounted gateway and bin-mounted sensor nodes revealed marked differences in signal propagation along and across the tree rows, confirming that distance alone cannot capture the behavior of sub-GHz links in such environments. By introducing a two-dimensional direction-dependent path-loss model, we demonstrated that accounting for the canopy thickness the link actually traverses across the orchard grid improves the precision of the model. Compared with classical isotropic baselines, ITU-R vegetation attenuation, and multi-wall approaches, the proposed model yields a more homogeneous error distribution and better reflects the physical structure of the orchard.
These results underscore the need for direction-sensitive propagation models adapted to agricultural plantation patterns. As farms increasingly rely on distributed sensing and automated workflows, capturing these geometric effects becomes essential for planning reliable wireless infrastructures and optimizing node placement, coverage, and energy consumption.

\section{Future Work}
Future research can extend the present study in several directions. First, incorporating seasonal variations such as canopy growth, pruning stages, and moisture levels, especially in the southern Italian environmental conditions, would allow the model to reflect temporal changes in different situations. 
% Second, integrating measurements from additional antenna heights and transceiver orientations may further clarify the contribution of ground reflections and canopy layering. 

Another direction is the combination of the proposed direction-dependent formulation with machine-learning techniques to refine parameter estimation while preserving interpretability. Finally, validating the model across different orchards and also different planting systems (HD/SHD hedgerows and irregular traditional layouts) will help determine its generality and support the development of deployment guidelines for large-scale agricultural IoT systems.

\section{Acknowledgment}

M. Rowhani thanks the Department of Telecommunications at AGH University of Krakow for hosting the research stay under the supervision of Prof. Katarzyna Kosek-Szott. The access to the experimental olive orchard used for field measurements was provided by the Department of Agricultural Sciences of the University of Palermo. This study received partial support from the SMOOL project, CUP G79J18000760007,  and from SEEDS s.r.l.
\bibliography{bib}
\EOD
\begin{IEEEbiography}
[{\includegraphics[width=1in,height=1.25in,clip,keepaspectratio]{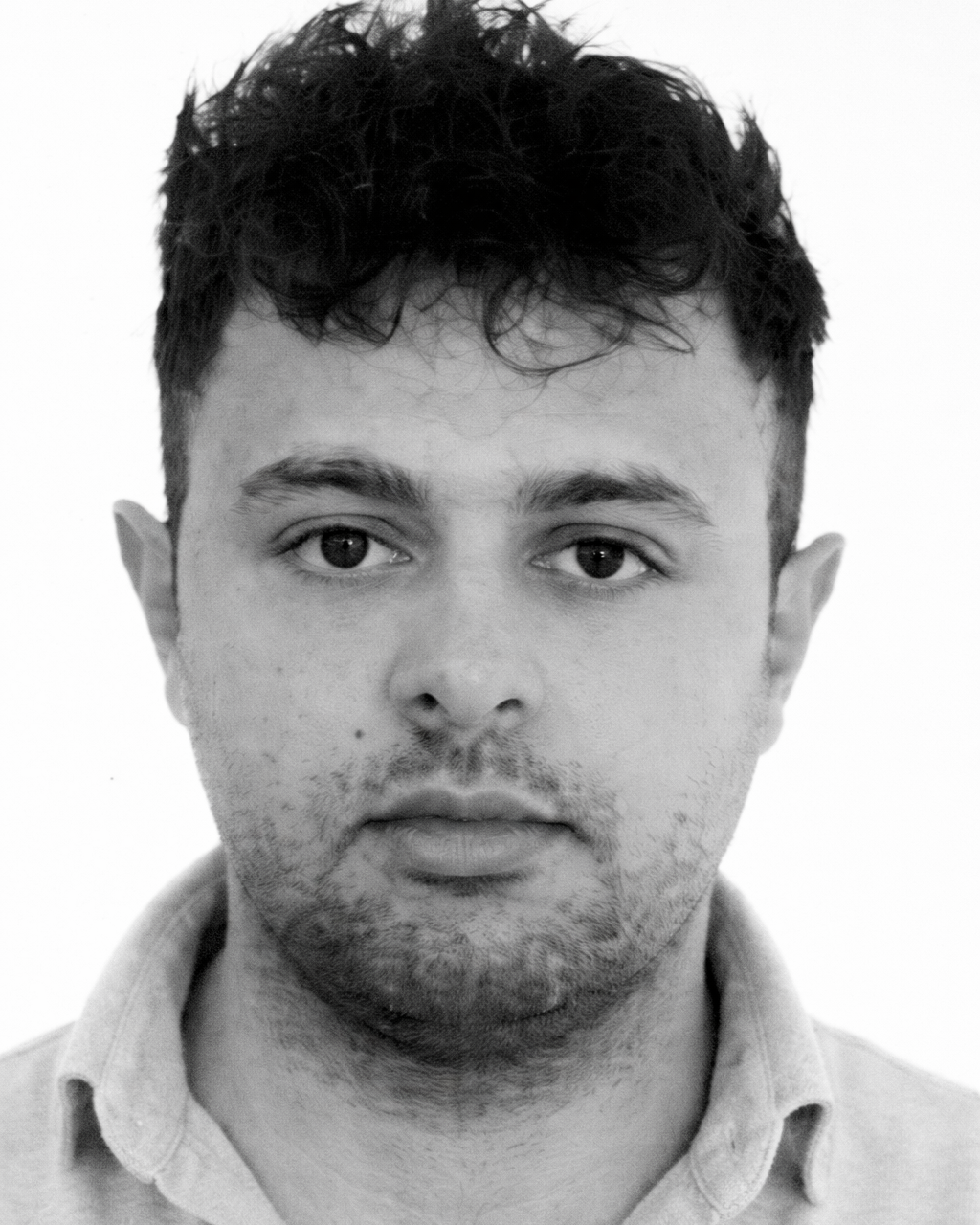}}]{Mohammad Rowhani Sistani}
He received the Ph.D. degree in Blockchain and Distributed
Ledger Technology (Agriculture and Agri-Food), jointly
awarded by the University of Camerino and the University
of Palermo, Italy, in 2026.

His research interests focus on blockchain-based traceability
systems, agri-food supply chains, smart agriculture, and
the integration of IoT and wireless communication
technologies in agricultural environments.

He has experience in agri-food export and applied research
on saffron traceability systems. His doctoral studies were financially supported by SEEDS s.r.l.
\end{IEEEbiography}

\begin{IEEEbiography}
[{\includegraphics[width=1in,height=1.25in,clip,keepaspectratio]{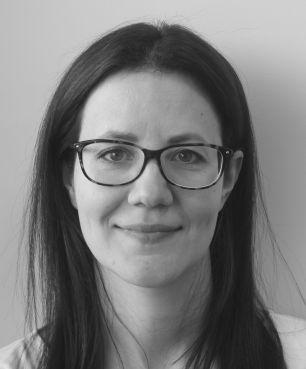}}]
{Katarzyna Kosek-Szott}
She received the Ph.D. degree in telecommunications in 2011
and the Habilitation degree in 2016. She is currently a Full Professor with
the Institute of Telecommunications, AGH University of Science and Technology,
Krakow, Poland. Her research focuses on wireless networking, including
wireless LANs, quality of service provisioning, novel amendments to the
IEEE~802.11 standard, 5G/6G networks, and coexistence of radio technologies.
She has coauthored more than 70 research papers and serves as a reviewer for
international journals and conferences. She has been involved in several
European research projects, including DAIDALOS~II, CONTENT, CARMEN, FLAVIA,
PROACTIVE, and RESCUE, as well as projects funded by the Polish government.
\end{IEEEbiography}

\begin{IEEEbiography}
[{\includegraphics[width=1in,height=1.25in,clip,]{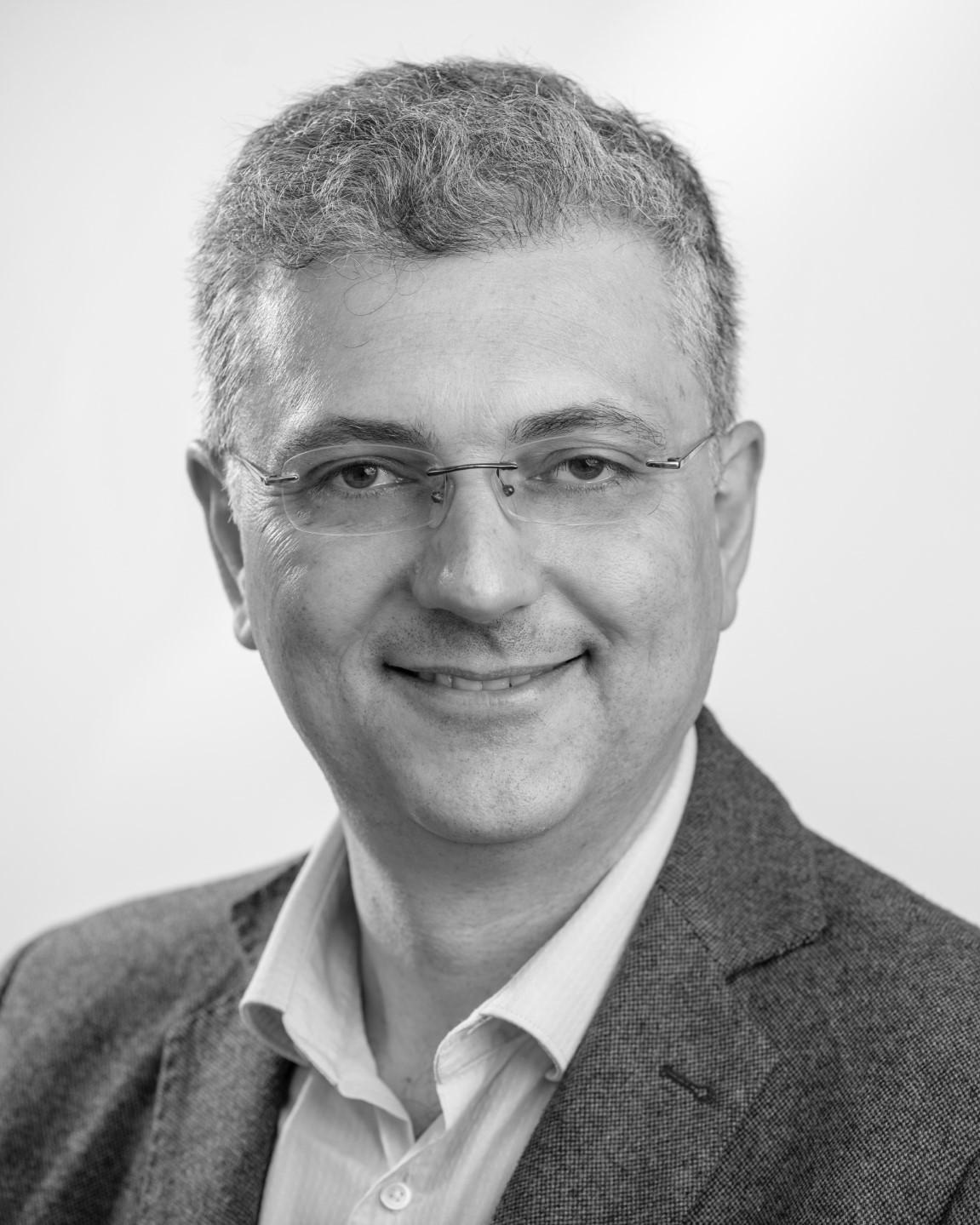}}]
{Pierluigi Gallo}
He received the Ph.D. degree in information engineering.
He is currently an Associate Professor with the Department of Engineering,
University of Palermo, Italy, and a researcher with CNIT.
His research interests include wireless networks, indoor localization,
communication security, blockchain technologies and architectures,
blockchain-based systems for secure data management and traceability,
and Internet of Things applications. He has been involved in several national and international
research projects related to wireless and IoT-enabled systems.
He is a Member of the IEEE.
\end{IEEEbiography}

\end{document}